\documentclass[reprint,superscriptaddress,amsmath,amssymb,aps,prb]{revtex4-2}

\usepackage{adjustbox}
\usepackage{placeins}
\usepackage{newtxtext}
\usepackage[varvw]{newtxmath}
\usepackage{graphicx}
\usepackage{dcolumn}
\usepackage{bm}
\usepackage{bbm}
\usepackage{csquotes}
\usepackage{mathtools}
\usepackage{physics}
\usepackage{subfigure}
\usepackage{tikz}

\usepackage{amsfonts,amsmath,amssymb,amsthm}
\usepackage[colorlinks=true,allcolors=blue,breaklinks=True]{hyperref}

\newcommand{\klr}[1]{\left(#1\right)}
\newcommand{\kle}[1]{\left[#1\right]}
\newcommand{\kls}[1]{\left|#1\right|}

\newcommand{\klsr}[1]{#1 \biggl|}
\newcommand{\kla}[1]{\langle#1\rangle}
\newcommand{\klg}[1]{\left\{#1\right\}}
\newcommand{\klgcases}[1]{\begin{cases}#1\end{cases}}

\newcommand{\varvec}[1]{\mathbf{#1}}
\renewcommand{\vv}[1]{\varvec{#1}}

\newcommand{\svec}[1]{\begin{pmatrix}#1\end{pmatrix}}
\newcommand{\matr}[1]{\svec{#1}}
\newcommand{\varvarvec}[1]{\bm{#1}}
\newcommand{\vvv}[1]{\varvarvec{#1}}



\newcommand{\mb}[1]{\textcolor{black}{#1}}

\begin{document}

\title{Functional approach to superfluid stiffness:\\
Role of quantum geometry in unconventional superconductivity}

\author{Maximilian Buthenhoff}
\email{buthenhoff.m.aa@m.titech.ac.jp}
\affiliation{Department of Physics, Institute of Science Tokyo, Ookayama, Meguro, Tokyo 152-8551, Japan}
\affiliation{Theoretical Physics III, Ruhr-University Bochum, D-44801 Bochum, Germany}

\author{Tobias Holder}
\email{tobiasholder@tauex.tau.ac.il}
\affiliation{School of Physics and Astronomy, Tel Aviv University, Tel Aviv, Israel}

\author{Michael M.\ Scherer}
\email{scherer@tp3.rub.de}
\affiliation{Theoretical Physics III, Ruhr-University Bochum, D-44801 Bochum, Germany}

\date{\today}

\begin{abstract}
Nontrivial quantum geometry of electronic bands has been argued to facilitate superconductivity even for the case of flat dispersions where the conventional contribution to the superfluid weight is suppressed by the large effective mass.
However, most previous work focused on the case of conventional superconductivity while many contemporary superconducting quantum materials are expected to host unconventional pairing.
Here, we derive a generalized expression for the superfluid weight employing mean-field BCS theory for systems with time-reversal symmetry in the normal state and arbitrary unconventional superconducting order with zero-momentum intraband pairing.
Our derivation reveals the necessity of incorporating functional derivatives of the grand potential with respect to the superconducting gap function. 
Through perturbative analysis in the isolated narrow-bands limit, we demonstrate that this contribution arises from quantum geometrical effects, specifically due to a nontrivial Wilczek-Zee connection. Utilizing the newly obtained expressions for the superfluid weight, we apply our framework to an extended Kane-Mele model, contrasting conventional $s$-wave superconductivity with chiral $d$-wave superconductivity.
\end{abstract}

\maketitle

\section{\label{sec:level1}Introduction}

While initial ideas on quantum geometry date back to the 1980s by Provost, Vallee \cite{provost1980riemannian}, and Berry \cite{berry1984quantal,berry1989quantum}, the systematic study of quantum geometry has recently been propelled into focus~\cite{yu2024quantum,Verma2025,hu2025geometric} due to a particularly tantalizing manifestation connected to the superfluid weight in flat bands~\cite{torma2023essay}. 
It is well known that the superfluid weight in a single-band Bardeen-Cooper-Schrieffer (BCS) theory (i.e., for a case of trivial quantum geometry) is only proportional to the inverse effective mass of the band~\cite{schrieffer1964theory},
\begin{align}
    D_{\mathrm{s}}^{\mathrm{conv}} \propto \frac{1}{m_{\mathrm{eff}}}\,.    
\end{align}
On the other hand, in a quantum material with many orbitals this estimate needs to be amended as the superfluid weight becomes a complicated function of the band structure and wave-function properties.
The possibility of a finite superfluid weight in the presence of dispersionless bands (where $m_{\mathrm{eff}}\to\infty$) was  recognized early by Khodel' and Shaginyan within the fermion-condensate scenario \cite{khodel1990superfluidity}. Later, Peotta and Törmä 
showed in Ref.~\cite{peotta2015superfluidity} with the help of multiband BCS theory that a nonzero superfluid weight can be obtained even in the presence of dispersionless bands, provided the system possesses
nontrivial quantum geometry.
In the wake of the experimental discovery of superconductivity in moir\'e systems which exhibit an unusually large single-particle effective mass~\cite{cao2018correlated,cao2018unconventional},
these ideas have garnered much attention, and have been sharpened and extended in several ways~\cite{liang2017band,hu2019geometric,julku2020superfluid,xie2020topology,torma2022superconductivity,huhtinen2022revisiting,torma2023essay,penttila2024flat,jiang2024geometric,lamponen2025superconductivity,yijian2025quantum}. 

For an isolated and nearly dispersionless (i.e., flat) band, the superfluid weight of a conventional superconducting order parameter has been determined and it was found that~\cite{huhtinen2022revisiting}
\begin{align}\label{eq:minimalmetric}
    D_{\mathrm{s}}^{\mathrm{geom}} \propto 
    \frac{|U|}{V}\sum_{\vv{k}}g^{\mathrm{min}}(\vv{k}) \,.
\end{align}
Here~$U$ is the (momentum-independent) interaction, and $g^{\mathrm{min}}(\vv{k})$ corresponds to a specific choice for the diagonal elements of the quantum metric 
$
    \hat{g}_{ij} =\mathrm{Re}[
    \hat{P}\partial_{k_i}\hat{P} \partial_{k_j}\hat{P}]
$, 
defined via the momentum overlap of the  projection $\hat{P}$ into the ground state.
The quantum metric~$g$ gives rise to an effective mass which is decoupled from the dispersive features of the band structure.
Indeed, the physical significance of~$g$ can be deduced purely from the real-space charge distribution~\cite{Sorella.Resta.1999,Martin.Souza.1999}, by association with the strength of dipole transition amplitude $Q_{ij}=\ev{\hat{r}_i(1-\hat P) \hat{r}_j}$ between ground state and excited states,
which is gauge invariant.
The quantum metric thus constitutes a characteristic length $\ell_g = \sqrt{\mathrm{tr}(g)}$ which is intrinsic to a quantum material~\cite{Verma2025}. 
However, unlike the quantum metric itself, the connection of~$g$ to the superfluid weight is less understood. 
Notably, as the index \enquote{min} indicates in  Eq.~\eqref{eq:minimalmetric}, the superfluid weight $D_{\mathrm{s}}$ does not depend on~$g$ directly, but instead requires a choice of an orbital embedding within the unit cell such that the trace of the quantum metric becomes minimal~\cite{huhtinen2022revisiting}.

Here, we elucidate the quantum geometric content contained in $D_{\mathrm{s}}$ by considering the superfluid weight of multiband BCS theory for an arbitrary 
unconventional superconducting order parameter with zero-momentum intraband pairing. 
To that end, 
electromagnetic interactions are included via a Peierls substitution~\cite{peierls1933theory}, leading to a substantially generalized estimate compared to Eq.~\eqref{eq:minimalmetric}. 
In technical terms, while in conventional BCS theory the gap function represents a single complex number for every fixed vector potential~$\vv{A}$, in unconventional BCS theory it represents a function with respect to the wave vector~$\vv{k}$~\cite{sigrist1991phenomenological}. Therefore, when calculating the superfluid weight for an unconventional superconducting state, one needs to take into consideration functional derivatives of the free energy with respect to the gap function. An analytical calculation of the functional derivative reveals that the associated contribution (which we dub \enquote{functional} superfluid weight) is driven by nontrivial quantum geometry beyond the quantum metric, and features Wilczek-Zee connections~\cite{wilczek1984appearance}.

We expect that this modification of the superfluid weight applies to many of the recently discovered van der Waals superconductors, for example in twisted bilayer graphene, twisted $\mathrm{WSe_2}$, and others, all of which have been argued to be of unconventional nature~\cite{xia2024superconductivity,oh2021evidence,tanaka2025superfluid}. We also note that similar considerations will be relevant regarding superconductivity derived from the Kohn-Luttinger mechanism~\cite{shavit2024quantum,jahin2024enhanced}.

\section{\label{sec:levelsum}Overview and main results}

We present a general expression for the superfluid weight of an unconventional superconducting order parameter in Eq.~\eqref{eq:superfluid-weight-bcs-trs}, which can be succinctly summarized as
\begin{align}
    D_{\mathrm{s}}=
    D_{\mathrm{s}}^{\mathrm{conv}} + D_{\mathrm{s}}^{\mathrm{geom}}-
    D_{\mathrm{s}}^{\mathrm{func}}\,.
\end{align}
Here $D_{\mathrm{s}}^{\mathrm{conv}}$ is the conventional contribution associated with the effective mass, $D_{\mathrm{s}}^{\mathrm{geom}}$ derives from the quantum metric, and $D_{\mathrm{s}}^{\mathrm{func}}$ is the functional superfluid weight which encodes information of the wave-function geometry beyond a single-momentum overlap. 
The explicit expression for the geometric superfluid weight is given in Eq.~\eqref{eq:superfluid-weight-geom}, while the functional superfluid weight is provided in Eq.~\eqref{eq:functional-superfluid-weight-wrt-skew-matrices}.

Another recently published work by Lamponen, Pöntys and Törmä also explores the superfluid weight of unconventional superconductors~\cite{lamponen2025superconductivity} with a focus on the case of nearest-neighbor pairing. 
Notably, Ref.~\cite{lamponen2025superconductivity} presents expressions corresponding to our key equations~\eqref{eq:superfluid-weight-bcs-trs},~\eqref{eq:superfluid-weight-geom}, and~\eqref{eq:functional-superfluid-weight-wrt-skew-matrices}. 
Wherever comparable, we find their results to be consistent with ours. Here, we have chosen to express our final equations for the superfluid weight by avoiding total derivatives of the gap function with respect to the vector potential. This is advantageous, for example, for numerical calculations.

Moreover, under the assumption of isolated narrow bands, we employ a perturbative approach to find the expressions in Eqs.~\eqref{eq:explicitgeom} and \eqref{eq:explicitfunc} for the geometric and functional superfluid weight, respectively.
While for conventional superconductivity it is well known that we can find a basis in which the functional contribution becomes zero, this is in general not the case for unconventional pairing mechanisms.
Most notably, we observe that 
$D_{\mathrm{s}}^{\mathrm{geom}}-D_{\mathrm{s}}^{\mathrm{func}}$ does not reduce to the minimal quantum metric according to Eq.~\eqref{eq:minimalmetric} unless additional conditions on the order parameter are imposed.

Specifically, in case of an isolated flat band, we find that
the functional superfluid weight $D_{\mathrm{s}}^{\mathrm{func}}$ can be related to the two-point fidelity magnitude $\zeta_{ij}(\vv{k},\vv{k}')$,
which can be expressed in terms of the Wilczek-Zee connection $e_{i,m}^{(n)}=i\braket{\psi_m}{\partial_{k_i}\psi_n}$~\cite{wilczek1984appearance} as
\begin{align}
    \zeta_{ij}(\vv{k},\vv{k}') = \sum_{m\neq n,m' \neq n'} \kls{e_{i,m}^{(n)}(\vv{k}) e_{j,m'}^{(n')}(\vv{k}')}\,.
\end{align}
For an unconventional superconducting state with pairing potential $U(\vv{k},\vv{k}')$ and order parameter $\Delta({\vv{k}})$, the essential ingredients to the functional superfluid weight thus become \mbox{$D_{\mathrm{s},ij}^{\mathrm{func}}\sim \sum_{\vv{k},\vv{k'}}M^{-1}({\vv{k},\vv{k'}}) \Delta(\vv{k}) \Delta({\vv{k'}})\zeta_{ij}(\vv{k},\vv{k'})$}, where $M^{-1}$ depends on the pairing interaction, the Bogoliubov coefficients, and the quasiparticle eigenvalues. However, note that this expression represents a bound only. In Eq.~\eqref{eq:exact-functional-contribution-isolated-band} we state the exact expression of the functional superfluid weight obtained through perturbation theory. In general, the functional contribution is not reducible to a single-momentum quantum geometry.

\begin{figure}
    \centering
    \includegraphics[width=\linewidth]{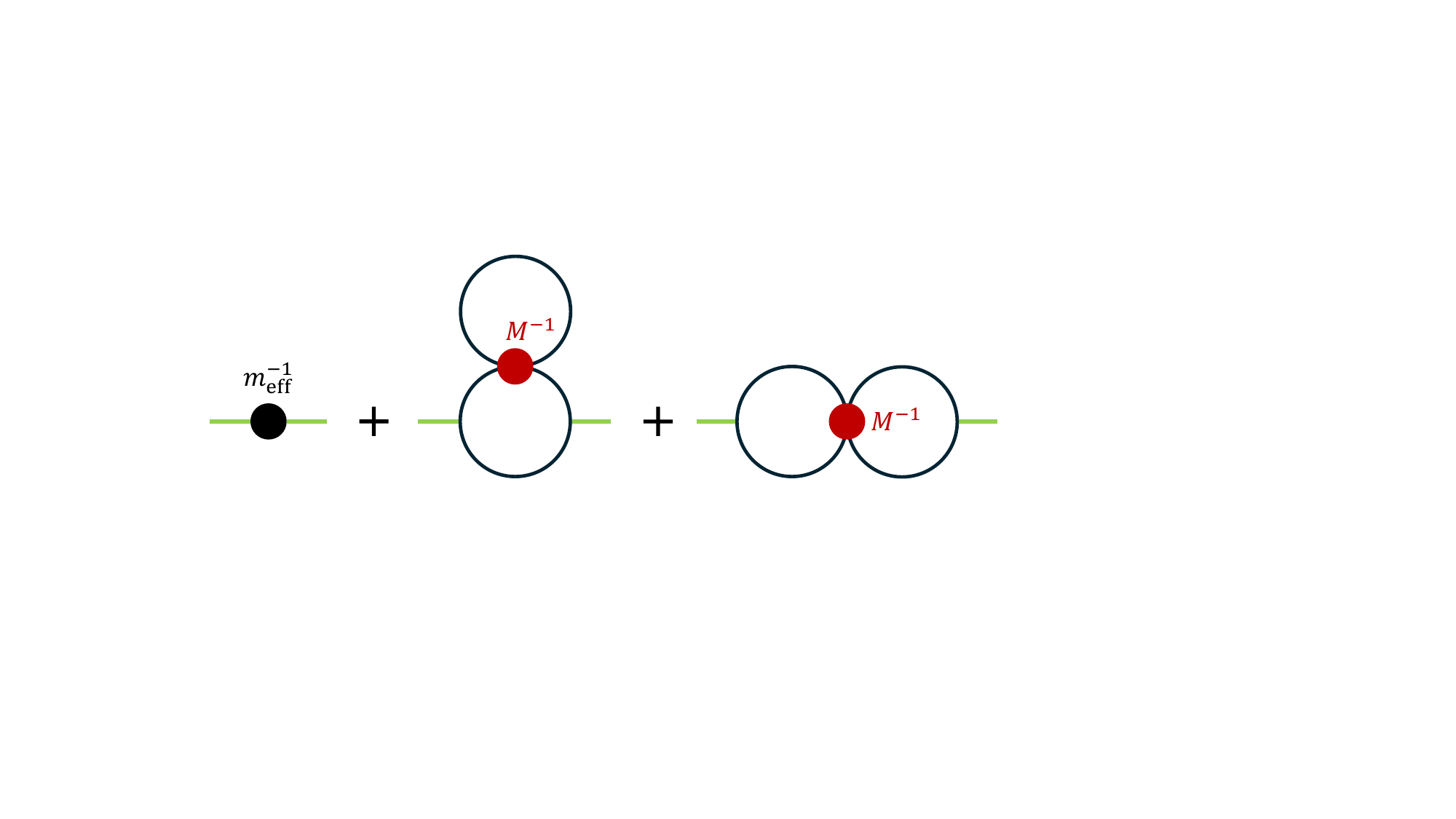}
    \caption{Principal structure of the second derivative of the free energy, comprising the tree-level contribution from the dispersion, and the two possible contractions of the pairing interaction, which contains information about the multiorbital nature of the band structure. Here, $M^{-1}$ depends on the pairing interaction, the Bogoliubov coefficients, and the quasiparticle eigenvalues.}
    \label{fig:schematics}
\end{figure}

This result can be understood by appealing to the multistate geometry~\cite{Bouhon2023,Avdoshkin2024,Antebi2024} that arises in interacting systems with several mutually independent momenta: The Wilczek-Zee connection is the natural generalization of the Berry connection for non-Abelian gauge fields. In a quantum material with many orbitals such contributions naturally arise due to interaction-mediated processes involving sums over frequency and momentum. 
The different contributions to the superfluid weight are depicted schematically in Fig.~\ref{fig:schematics}.
Since $D_{\mathrm{s}}$ is the second-order derivative in the free energy, one can expect at least two independent momenta to enter the final estimate, unless the interaction and order parameter are both independent of momentum.

To illustrate the different contributions from a numerical point of view we investigate an extended version of the Kane-Mele model~\cite{kane2005quantum,kane2005z,lau2022universal} that includes hopping terms up to fourth-nearest neighbors. 
This model has a rich topological phase diagram with a topologically nontrivial flat band configuration and it preserves time-reversal symmetry. 
Moreover, the standard Kane-Mele model has been proposed as a toy model for twisted $\mathrm{WSe_2}$~\cite{mak2022semiconductor}
which was recently observed to host superconductivity~\cite{xia2024superconductivity,Guo_2025}.
For comparison, we discuss the conventional $s$-wave and unconventional chiral $d$-wave superconducting state in this model.
For the cases we tested, $D_{\mathrm{s}}^{\mathrm{func}}$ introduces finite but small modifications of the total superfluid weight, which 
suggests that Eq.~\eqref{eq:minimalmetric} is already a viable first estimate in many practical situations. We elucidate and quantify deviations in detail in this work. 

\section{\label{sec:bcs-theory}Mean-field BCS theory for arbitrary effective attractive pairing potential}

Although BCS theory was initially developed based on Cooper pairing mediated by lattice vibrations~\cite{cooper1956bound,bardeen1957theory}, it has been shown that quasiparticle properties can also describe unconventional superconductors~\cite{zhu2016bogoliubov,sigrist1991phenomenological}. In the following, we do not focus on any specific pairing mechanism but instead assume the existence of an effective mechanism responsible for the attractive two-particle interaction.

\subsection{Superconducting mean-field Hamiltonian for time-reversal symmetric single-particle systems}\label{sec:bcs-theory-setup}

For an arbitrary attractive effective two-particle interaction $U(\vv{k},\vv{k}')$ the BCS partition function is given by~\cite{sigrist1991phenomenological,weinberg1996quantum}
\begin{align}
    Z(\vv{q}) =& \int \mathcal{D}[\psi,\psi^\dagger] \exp\!\Bigg[-\sum_{\vv{k}} \psi_{\vv{k}\alpha}^\dagger H_{\alpha\beta}(\vv{k} - \vv{q}) \psi_{\vv{k}\beta} \nonumber \\
    &+ \frac{1}{2V} \sum_{\vv{k},\vv{k}'} U(\vv{k},\vv{k}') \psi^\dagger_{\vv{k}\alpha} \psi^\dagger_{-\vv{k}\beta} \psi_{-\vv{k}'\beta} \psi_{\vv{k}'\alpha}\Bigg]\,.
    \label{eq:bcs-partition-function}
\end{align}
Here, we have used the Einstein sum convention and normalized the measure accordingly. 
Further, $V$ represents the volume of the Brillouin zone, $\psi^\dagger_{\vv{k}\alpha}$ and $\psi_{\vv{k}\alpha}$ are creation and annihilation operators of an electron in orbital $\alpha = 1,\hdots,N_{\mathrm{B}}$ with momentum $\vv{k}$, and $H_{\alpha\beta}(\vv{k})$ is the single-particle Hamiltonian in momentum space. 
Furthermore, with the help of the Peierls substitution we have included electromagnetic interactions, i.e., $\vv{q} = \vv{A}$ (with convention $e = 1$)~\cite{peierls1933theory}.

We perform a Hubbard-Stratonovich transformation in the pairing channel introducing auxiliary bosonic fields $\Delta_{\alpha\beta}(\vv{k})$ to obtain the Hubbard-Stratonovich path integral with action
\begin{align}
    &S_{\mathrm{HS}}[\vv{q}; \Delta,\Delta^\dagger] = \frac{V}{2} \sum_{\vv{k},\vv{k}'} U^{-1}(\vv{k},\vv{k}') \Delta_{\alpha\beta}^\dagger(\vv{k}) \Delta_{\beta\alpha}(\vv{k}') \nonumber \\
    &\quad - \ln\klr{\int \mathcal{D}[\psi,\psi^\dagger] \exp\!\klr{-S[\vv{q}; \psi,\psi^\dagger,\Delta,\Delta^\dagger]}} \,,
    \label{eq:hubbard-stratonovich-action}
\end{align}
where
\begin{align}
    &S[\vv{q}; \psi,\psi^\dagger,\Delta,\Delta^\dagger] = \sum_{\vv{k}} \psi_{\vv{k}\alpha}^\dagger H_{\alpha\beta}(\vv{k} - \vv{q}) \psi_{\vv{k}\beta} \nonumber \\
    &\quad + \frac{1}{2} \sum_{\vv{k}} \Delta^\dagger_{\alpha\beta}(\vv{k}) \psi_{-\vv{k}\beta} \psi_{\vv{k}\alpha} + \frac{1}{2} \sum_{\vv{k}} \Delta_{\beta\alpha}(\vv{k}) \psi^\dagger_{\vv{k}\alpha} \psi^\dagger_{-\vv{k}\beta}\,,
    \label{eq:action-mean-field-electrons}
\end{align}
represents the fermionic part of the action. 
To find the field configurations for the auxiliary fields that minimize the Hubbard-Stratonovich action, we carry out a saddle-point approximation, yielding~\cite{weinberg1996quantum,sigrist1991phenomenological}
\begin{align}
    \Delta_{\alpha\beta}(\vv{q}; \vv{k}) = \frac{1}{V} \sum_{\vv{k}'} U(\vv{k},\vv{k}') \kla{\psi_{-\vv{k}'\alpha} \psi_{\vv{k}'\beta}}_{\vv{q}} \,,
    \label{eq:self-consistent-equation}
\end{align}
where $\kla{\,\cdot\,}_{\vv{q}}$ denotes the average with respect to the Hubbard-Stratonovich path integral. 
Since the single-particle Hamiltonian is $\vv{q}$ dependent, the path integral in the logarithm is $\vv{q}$ dependent. 
Therefore, the values of the variables $\Delta$ and $\Delta^\dagger$ that minimize the Hubbard-Stratonovich action, are also implicitly $\vv{q}$ dependent. 
To emphasize the $\vv{q}$ dependence of the auxiliary fields we have indicated this explicitly in the self-consistent equation~\eqref{eq:self-consistent-equation}. 
Note that for every fixed $\vv{q}$ the auxiliary field represents a $\vv{k}$-dependent function, called gap function. Only if we assume $s$-wave pairing symmetry, i.e., a constant attractive interaction strength $U_0$, the gap function $\Delta$ is a complex number independent of $\vv{k}$ for every fixed $\vv{q}$. 

Let us assume that our single-particle Hamiltonian describing the normal state possesses time-reversal symmetry (TRS). 
The gap function may break TRS, however. 
Then, the Bogolioubov-de Gennes (BdG) Hamiltonian reads as~\cite{bogoliubov1958new,de1966superconductivity}
\begin{align}
    \mathcal{H}_{\mathrm{BdG}}(\vv{k},\vv{q}) \coloneqq \matr{H(\vv{k}-\vv{q}) - \mu \mathbbm{1} & \Delta(\vv{k})\\ \Delta^\dagger(\vv{k}) & -H(\vv{k}+\vv{q}) + \mu \mathbbm{1}}\,,
    \label{eq:bdg-hamiltonian}
\end{align}
which represents a ($2N_{\mathrm{B}} \times 2N_{\mathrm{B}}$) matrix. Using this matrix and with the help of the fermionic anticommutation relations we can write the mean-field Hamiltonian as
\begin{align}
    H_{\mathrm{MF}}(\vv{q}) &= \frac{1}{2} \sum_{\vv{k}} \Psi_{\vv{k}}^\dagger \mathcal{H}_{\mathrm{BdG}}(\vv{k},\vv{q}) \Psi_{\vv{k}} \nonumber \\
    &\quad + \frac{1}{2} \sum_{\vv{k}} \tr(H(\vv{k}-\vv{q}) - \mu \mathbbm{1}) \nonumber \\
    &\quad + \frac{V}{2} \sum_{\vv{k},\vv{k}'} U^{-1}(\vv{k},\vv{k}') \Delta_{\alpha\beta}^\dagger(\vv{k}) \Delta_{\beta\alpha}(\vv{k}')\,,
    \label{eq:mean-field-hamiltonian-for-nambu}
\end{align}
where $\Psi_{\vv{k}} = (\psi_{\vv{k}1},\psi_{\vv{k}2},\hdots,\psi^\dagger_{-\vv{k}1},\psi^\dagger_{-\vv{k}2},\hdots)$ is the Nambu spinor. We diagonalize $H(\vv{k})$ by a matrix $S(\vv{k})$ as
\begin{align}
    \varepsilon_{\vv{k}} = S^\dagger(\vv{k})H(\vv{k})S(\vv{k})\,,
    \label{eq:diagonalization-single-particle}
\end{align}
where $\varepsilon_\vv{k}$ is a diagonal matrix. 
Then, we introduce new creation and annihilation operators by
\begin{align}
    \svec{d_{\vv{k}}(\vv{q})\\ e_{\vv{k}}(\vv{q})} \coloneqq \matr{S^\dagger(\vv{k}-\vv{q}) & \\ & S^\dagger(\vv{k}+\vv{q})} \Psi_{\vv{k}} \,.
    \label{eq:basis-transformation-dk}
\end{align}
In this basis, the BdG Hamiltonian becomes single-particle diagonalized and can be represented as 
\begin{align}
    \mathcal{H}_{\vv{k}}(\vv{q}) \coloneqq \matr{\varepsilon_{\vv{k}-\vv{q}} - \mu\mathbbm{1} & \mathcal{D}_k(\vv{q})\\ \mathcal{D}^\dagger_k(\vv{q}) & \varepsilon_{\vv{k}+\vv{q}} + \mu\mathbbm{1}}\,,
    \label{eq:diagonalized-bdg-hamiltonian}
\end{align}
where the off-diagonal block matrices are given by
\begin{align}
    \mathcal{D}_k(\vv{q}) \coloneqq S^\dagger(\vv{k}-\vv{q}) \Delta(\vv{k}) S(\vv{k}+\vv{q}) \,,
\end{align} 
which are in general dependent of the normalization phase factor used in the matrix $S(\vv{k})$. It is useful to work in this basis to obtain expressions for the superfluid weight within the isolated narrow-bands limit (cf. Sec.~\ref{sec:isolated-band-limit}). 

Let us further assume that the single-particle diagonalized BdG Hamiltonian~\eqref{eq:diagonalized-bdg-hamiltonian} can be diagonalized by a $(2N_{\mathrm{B}} \times 2N_{\mathrm{B}})$ matrix $W_{\vv{k}}(\vv{q})$ such that
\begin{align}
    E_{\vv{k}}(\vv{q}) = W_{\vv{k}}^\dagger(\vv{q}) \mathcal{H}_{\vv{k}}(\vv{q}) W_{\vv{k}}(\vv{q})
    \label{eq:diagonalization-bdg-hamiltonian}
\end{align}
is a diagonal matrix with eigenvalues $E_{\vv{k}n}(\vv{q})$. At $\vv{q} = 0$, the eigenvalues $E_{\vv{k}n} \equiv E_{\vv{k}n}(0)$ are called Bogoliubov eigenvalues or quasiparticle eigenvalues of the BdG Hamiltonian, and the eigenvectors \cite{sigrist1991phenomenological}
\begin{align}
    W_{\vv{k}} \equiv W_{\vv{k}}(0) \coloneqq \matr{\mathcal{U}_{\vv{k}} & \mathcal{V}_{\vv{k}}\\ \mathcal{V}_{-\vv{k}}^\ast & \mathcal{U}_{-\vv{k}}^\ast} 
    \label{eq:bogliubov-coefficients-def}
\end{align}
are called Bogoliubov coefficients. Due to the unitarity of $W_{\vv{k}}$, the Bogoliubov coefficients satisfy the relations
\begin{align}
    \klgcases{
        \mathcal{U}_{\vv{k}} \mathcal{U}_{\vv{k}}^\dagger + \mathcal{V}_{\vv{k}} \mathcal{V}_{\vv{k}}^\dagger = \mathbbm{1}\,,\\
        \mathcal{U}_{\vv{k}} \mathcal{V}^T_{-\vv{k}} + \mathcal{V}_{\vv{k}} \mathcal{U}_{-\vv{k}}^T = 0 \,.
    }
   \label{eq:relations-bog-coeffs-trs}
\end{align}
Since the BdG Hamiltonian satisfies $\mathcal{C} \mathcal{H}_{\mathrm{BdG}}(\vv{k}) \mathcal{C}^{-1} = -\mathcal{H}_{\mathrm{BdG}}(-\vv{k})$, where $\mathcal{C} = \sigma_x \mathcal{K}$ with $\mathcal{K}$ complex conjugation, every eigenvalue at $\vv{k}$ has a partner at $-\vv{k}$ with opposite energy such that the we have the relation $E_{\vv{k}n} = -E_{-\vv{k}(n+N_{\mathrm{B}})}$~\cite{sigrist1991phenomenological,zirnbauer2021particle}. These eigenvalues determine the grand potential. Since
\begin{align}
    -T\ln(1 + e^{-E/T}) \xrightarrow{T \to 0} \klgcases{E & E < 0\\ 0 & E \ge 0}
\end{align}
the grand potential at zero temperature is given by
\begin{align}
    \Omega(\vv{q}) &= - \frac{1}{4} \sum_{\vv{k},n}|E_{\vv{k}n}(\vv{q})| + \frac{1}{2} \sum_{\vv{k}} \tr(\varepsilon_{\vv{k}-\vv{q}} - \mu\mathbbm{1}) \nonumber \\
    &\quad + \frac{V}{2} \sum_{\vv{k},\vv{k}'} U^{-1}(\vv{k},\vv{k}') \Delta_{\alpha\beta}^\dagger(\vv{k}) \Delta_{\beta\alpha}(\vv{k}') \,.
    \label{eq:grand-potential-mean-field-bcs}
\end{align}
A generalization to nonzero temperature is straightforward and modifies the result accordingly.

\subsection{General expression for the superfluid weight}\label{sec:calculation-superfluid-weight-functional}

\subsubsection{Derivation}

The superfluid weight is defined as \cite{taylor2006pairing}
\begin{align}
    D_{\mathrm{s},ij} = \frac{1}{V} \frac{\mathrm{d}^2 F}{\mathrm{d}q_i \mathrm{d}q_j} \klsr{}_{\vv{q} = 0} \,.
\end{align}
Hence, we need to determine the total derivative of the free energy $F = \Omega + \mu N$.
To that end, we follow the steps given in Ref.\ \cite{huhtinen2022revisiting} and generalize the formulas accordingly.

Let us assume that $N = -\partial \Omega / \partial \mu$ is constant as a function in~$\vv{q}$. 
Since the grand potential $\Omega$ given in Eq.~\eqref{eq:grand-potential-mean-field-bcs} is a function in~$\vv{q}$ and $\mu \equiv \mu(\vv{q})$, and a functional in $\Delta_{\alpha\beta} \equiv \Delta_{\alpha\beta}(\vv{q}; \cdot)$ for $\alpha,\beta = 1,\hdots,N_{\mathrm{B}}$, according to the chain rule of differentiation the first derivative can be expressed as
\begin{widetext}
\begin{align}
    \frac{\mathrm{d}\Omega}{\mathrm{d}q_i} =& \frac{\partial \Omega}{\partial q_i} + \frac{\partial \Omega}{\partial \mu} \frac{\mathrm{d}\mu}{\mathrm{d}q_i} + \sum_{\vv{k},\alpha,\beta} \Bigg(\frac{\delta \Omega[\Delta_{\alpha\beta}^{\mathrm{R}}(\vv{q})]}{\delta\Delta_{\alpha\beta}^{\mathrm{R}}(\vv{q};\vv{k})} \frac{\mathrm{d}\Delta_{\alpha\beta}^{\mathrm{R}}(\vv{q};\vv{k})}{\mathrm{d}q_i} 
    + \frac{\delta \Omega[\Delta_{\alpha\beta}^{\mathrm{I}}(\vv{q})]}{\delta\Delta_{\alpha\beta}^{\mathrm{I}}(\vv{q};\vv{k})} \frac{\mathrm{d}\Delta_{\alpha\beta}^{\mathrm{I}}(\vv{q};\vv{k})}{\mathrm{d}q_i} \Bigg) \,, 
    \label{eq:total-derivative-grandpotential}
\end{align}
where we expand the grand potential as a Taylor series around $\Delta_{\alpha\beta}^{\mathrm{R,I}}(\vv{q})$ and use the formula given in Ref.~\cite{ernzerhof1994taylor}. Here, we set $\Delta_{\alpha\beta}^{\mathrm{R}} \equiv \mathrm{Re}(\Delta_{\alpha\beta})$ and $\Delta_{\alpha\beta}^{\mathrm{I}} \equiv \mathrm{Im}(\Delta_{\alpha\beta})$. Eq.~\eqref{eq:total-derivative-grandpotential} is a natural generalization for unconventional superconducting states of the expression for the total derivative provided by Huhtinen \textit{et al.}\ in Ref.~\cite{huhtinen2022revisiting}. In total, we obtain due to TRS of the single-particle Hamiltonian
\begin{align}
    D_{\mathrm{s},ij} = \frac{1}{V} \frac{\partial^2 \Omega}{\partial q_i \partial q_j} \klsr{}_{\vv{q} = 0} - \underbrace{\frac{1}{V} \sum_{\mu,\nu = \mathrm{R},\mathrm{I}} \sum_{\vv{k},\vv{k}'} \frac{\delta}{\delta \Delta_{\alpha_1\beta_1}^\mu(\vv{q}; \vv{k})}  \klr{\frac{\partial \Omega}{\partial q_i}} \kle{\frac{\delta^2 \Omega}{\delta \Delta^\mu_{\alpha_1\beta_1}(\vv{q};\vv{k}) \delta \Delta^\nu_{\alpha_2\beta_2}(\vv{q};\vv{k}')}}^{-1} \frac{\delta}{\delta \Delta_{\alpha_2\beta_2}^\nu(\vv{q}; \vv{k}')}  \klr{\frac{\partial \Omega}{\partial q_j}} \klsr{}_{\vv{q} = 0}}_{\eqqcolon D_{\mathrm{s},ij}^{\mathrm{func.}}} \,.
    \label{eq:superfluid-weight-bcs-trs}
\end{align}
\end{widetext}
Note that if the gap function is independent of $\vv{q}$, the components of the superfluid weight $D_{\mathrm{s},ij}$ are completely determined by the second partial derivative of the grand potential $\Omega$ with respect to $q_i$ and $q_j$ and we do not need to worry about functional derivatives at all. But in general this contribution is nonzero and to distinguish the first contribution (that contains the conventional and geometrical contribution to the superfluid weight) from the second one that contains functional derivatives, we also call it \enquote{functional superfluid weight} and abbreviate it by $D_{\mathrm{s},ij}^{\mathrm{func.}}$. 

Sometimes (e.g., in Ref.\ \cite{huhtinen2022revisiting}) this contribution is included in the geometrical contribution. However, here we want to discuss this contribution separately as its physical content is of a slightly different geometrical nature which is not the quantum metric, but rather due to generalized quantum geometry. This becomes evident in the isolated narrow-bands limit we discuss in the following Sec.~\ref{sec:isolated-band-limit}. 

\subsubsection{Conventional and geometrical contribution}

Let us take a look at the first term in Eq.~\eqref{eq:superfluid-weight-bcs-trs}. 
This contribution is identical to the one present in standard BCS theory for conventional superconductivity. 
It contains the conventional and the geometrical contributions to the superfluid weight
\begin{align}
    \frac{1}{V} \frac{\partial^2 \Omega}{\partial q_i \partial q_j} \klsr{}_{\vv{q} = 0} = D_{\mathrm{s},ij}^{\mathrm{conv}} + D_{\mathrm{s},ij}^{\mathrm{geom}} \,.
    \label{eq:conv-geom-sw}
\end{align}
Analytical expressions for this quantity are derived in Ref.~\cite{peotta2015superfluidity} for systems with TRS and in the Supplemental Material of Ref.~\cite{xie2020topology} for systems breaking TRS.

If we assume that the single-particle Hamiltonian has TRS, the conventional contribution depends on the curvature of the energy bands and is provided by
\begin{align}
    D_{\mathrm{s},ij}^{\text{conv}} = \frac{1}{2V} \sum_{\vv{k}} \tr[(\mathcal{V}_{\vv{k}} \mathcal{V}_{\vv{k}}^\dagger + \mathcal{V}_{-\vv{k}}^\ast \mathcal{V}_{-\vv{k}}^T) \partial_{k_i}\partial_{k_j} \varepsilon_{\vv{k}}]\,.
    \label{eq:superfluid-weight-conv-def}
\end{align}
The geometrical contribution is given by
\begin{align}
    D_{\mathrm{s},ij}^{\mathrm{geom}} =& -\frac{1}{4V} \sum_{\vv{k},n} \mathrm{sgn}(E_{\vv{k}n}) \partial_{q_i}\partial_{q_j} E_{\vv{k}n}(\vv{q})\klsr{}_{\vv{q}=0} \nonumber \\
    &+ \frac{1}{2V} \sum_{\vv{k}} \tr(\partial_{k_i} \partial_{k_j} \varepsilon_{\vv{k}}) - D_{\mathrm{s},ij}^{\mathrm{conv}}\,.
    \label{eq:superfluid-weight-geom}
\end{align}
It is also possible to use the Hellman-Feynman theorem (cf.\ Appendix~\ref{app:hellmann-feynman}) to express the second derivative of the quasiparticle eigenvalues in terms of the Bogoliubov coefficients. Within the isolated-bands limit it becomes evident why this contribution is called the geometrical contribution (see below).
%

\subsubsection{Functional contribution for intraband pairing}

Next, we bring the functional superfluid weight defined in Eq.~\eqref{eq:superfluid-weight-bcs-trs} into a usable shape. 
From now on, we assume intraband pairing only, i.e., we assume that the gap function is diagonal in each band $\Delta = \mathrm{diag}(\Delta_1, \hdots,\Delta_{N_{\mathrm{B}}})$. An inclusion of interband pairing gives rise to different contributions which have been discussed, for example, in Refs.~\cite{yijian2025quantum,lamponen2025superconductivity} in more detail. See also Ref.~\cite{putzer2025eliashberg}, where the impact of band-off-diagonal pairing on the superfluid stiffness in twisted graphene systems has been discussed. As in Sec.~\ref{sec:bcs-theory-setup}, the $2N_{\mathrm{B}}$ eigenvectors are given by the columns of the matrix
\begin{align}
    V_{\vv{k}}(\vv{q}) \coloneqq \matr{S(\vv{k}-\vv{q}) & 0 \\ 0 & S(\vv{k}+\vv{q})} \cdot W_{\vv{k}}(\vv{q})\,.
\end{align}
\indent First of all we would like to calculate the functional derivative of $\partial \Omega/ \partial q_i$ with respect to $\Delta^\mu_\alpha(\vv{q}; \vv{k}')$ at $\vv{q} = 0$. Due to the Hellman-Feynman theorem it is given by
\begin{align}
    &\frac{\delta}{\delta \Delta_\alpha^\mu(\vv{k}')} \klr{\frac{\partial \Omega}{\partial q_i}} \klsr{}_{\vv{q} = 0} \label{eq:functional-derivative-wrt-Delta} \\
    &\quad = -\frac{1}{4} \sum_{\vv{k},n} \mathrm{sgn}(E_{\vv{k}n})\, \klr{G^{\mathrm{(1)}}_{\vv{k}n, \alpha i, \mu}(\vv{k}') + G^{\mathrm{(2)}}_{\vv{k}n, \alpha i, \mu}(\vv{k}')}\,, \nonumber
\end{align}
with
\begin{align}
    G^{\mathrm{(1)}}_{\vv{k}n, \alpha i, \mu}(\vv{k}') &= \kle{V_{\vv{k}}^\dagger \delta_{\Delta^\mu_\alpha(\vv{k}')} \partial_{q_i} \mathcal{H}_{\mathrm{BdG}}(\vv{k},\vv{q}) |_{\vv{q} = 0} V_{\vv{k}}}_{nn}\,,\\
    G^{\mathrm{(2)}}_{\vv{k}n, \alpha i, \mu}(\vv{k}') \!&=\!\! \sum_{m \neq n}\! \klr{\frac{\kle{r_{\alpha,\mu}(\vv{k},\vv{k}')}_{nm} \kle{\tilde{r}_{i}(\vv{k})}_{mn}}{E_{\vv{k}n} \!-\! E_{\vv{k}m}} - (m\! \leftrightarrow\! n)},
\end{align}
where 
\begin{align}
    r_{\alpha,\mu}(\vv{k},\vv{k}') &= V_{\vv{k}}^\dagger \delta_{\Delta^\mu_\alpha(\vv{k}')} \mathcal{H}_{\mathrm{BdG}}(\vv{k},\vv{q}) |_{\vv{q}=0} V_{\vv{k}}\,, \\
    \tilde{r}_{i}(\vv{k}) &= V_{\vv{k}}^\dagger \partial_{q_i} \mathcal{H}_{\mathrm{BdG}}(\vv{k},\vv{q}) |_{\vv{q}=0} V_{\vv{k}} \,.
\end{align}
First, note that $G^{\mathrm{(1)}}_{\vv{k}n, \alpha i,\mu}(\vv{k}') = 0$ as the functional derivative of the diagonal block matrices is zero (the single-particle Hamiltonian does not depend on the gap function) and the partial derivative of the off-diagonal block matrices with respect to~$q_i$ is zero (the gap function depends only implicitly on~$\vv{q}$). 
To calculate the second contribution, we need the matrix elements of the two matrices that occur. Let $(\mathbbm{1}_{\alpha})_{\beta\gamma} \coloneqq \delta_{\alpha\beta} \delta_{\alpha\gamma}$ and define 
\begin{align}
    &A_{\alpha,\mu}(\vv{k}) \nonumber \\
    &= \klgcases{
        \mathcal{U}_{\vv{k}}^\dagger S^\dagger(\vv{k}) \mathbbm{1}_\alpha S(\vv{k}) \mathcal{U}^\ast_{-\vv{k}} + \mathcal{V}_{-\vv{k}}^T S^\dagger(\vv{k}) \mathbbm{1}_\alpha S(\vv{k}) \mathcal{V}_{\vv{k}} & \mu = \mathrm{R} \\
        i\klr{\mathcal{U}_{\vv{k}}^\dagger S^\dagger(\vv{k}) \mathbbm{1}_\alpha S(\vv{k}) \mathcal{U}^\ast_{-\vv{k}} - \mathcal{V}_{-\vv{k}}^T S^\dagger(\vv{k}) \mathbbm{1}_\alpha S(\vv{k}) \mathcal{V}_{\vv{k}}} & \mu = \mathrm{I}
    }
    \label{eq:definition-matrix-A}
\end{align}
and
\begin{align}
    B_i(\vv{k}) =& \mathcal{U}_{\vv{k}}^\dagger S^\dagger(\vv{k}) \partial_{k_i} H(\vv{k}) S(\vv{k}) \mathcal{V}_{\vv{k}} \nonumber \\ 
    &+ \mathcal{V}_{-\vv{k}}^T S^\dagger(\vv{k}) \partial_{k_i} H(\vv{k}) S(\vv{k}) \mathcal{U}^\ast_{-\vv{k}} \,.
\end{align}
Then, the necessary matrices are given by
\begin{align}
    \kle{r_{\alpha,\mu}(\vv{k},\vv{k}')}_{mn} &=  \kle{\matr{\ast & A_{\alpha,\mu}(\vv{k}') \\ A^\dagger_{\alpha,\mu}(\vv{k}') & \ast}}_{mn} \delta_{\vv{k},\vv{k}'} 
    \label{eq:one-func-derivative-with-wks} 
\end{align}
and
\begin{align}
    \kle{\tilde{r}_{i}(\vv{k})}_{mn}= \kle{\matr{\ast & B_i(\vv{k}) \\ B_i^\dagger(\vv{k}) & \ast }}_{mn}\,,
\end{align}
where $\ast$ denotes the diagonal elements that cancel out due to the $\mathrm{sgn}$ function. Using these results and the matrix summation formulas given in Appendix~\ref{sec:summation} [cf.\ Eq.~\eqref{eq:important-sum-fh}], we determine the functional derivative in~\eqref{eq:functional-derivative-wrt-Delta} to be 
\begin{align}
    &\frac{\delta}{\delta \Delta_\alpha^\mu(\vv{k}')} \klr{\frac{\partial \Omega}{\partial q_i}} \klsr{}_{\vv{q} = 0} = -\sum_{n,m=1}^{N_{\mathrm{B}}} \frac{\mathrm{Re}[(A_{\alpha,\mu}(\vv{k}'))_{nm} (B_i(\vv{k}'))^\ast_{nm}]}{E_{\vv{k}'n} + E_{-\vv{k}'m}}\,.
\end{align}
\indent Next we need to care about the second functional derivative of the grand potential at zero temperature which is given in Eq.~\eqref{eq:grand-potential-mean-field-bcs}. 
Employing the Feynman-Hellman theorem, again, the second functional derivative is found to be
\begin{align}
    &\frac{\delta^2 \Omega}{\delta \Delta^\mu_\alpha(\vv{q};\vv{k}_1) 
    \delta \Delta^\nu_\beta(\vv{q};\vv{k}_2)} \klsr{}_{\vv{q}=0} = V U^{-1}(\vv{k}_1,\vv{k}_2) \delta_{\alpha\beta} \delta_{\mu\nu} \nonumber \\
    &-\frac{1}{4} \sum_{\substack{\vv{k},n,m \\ m \neq n}} \frac{\mathrm{sgn}(E_{\vv{k}n})}{E_{\vv{k}n} - E_{\vv{k}m}} 
    \Big( [r_{\alpha,\mu}(\vv{k},\vv{k}_1)]_{nm} [r_{\beta,\nu}(\vv{k},\vv{k}_2)]_{mn} \nonumber \\
    &\quad+ ((\alpha,\mu,\vv{k}_1) \leftrightarrow (\beta,\nu,\vv{k}_2))
    \Big)  \,,
\end{align}
i.e., the second functional derivative depends on the matrices $A_{\alpha,\mu}(\vv{k}_1)$ and $A_{\beta,\nu}(\vv{k}_2)$ defined in Eq.~\eqref{eq:definition-matrix-A}. We use Eq.~\eqref{eq:definition-sum-s} to find that
\begin{align}
    &\frac{\delta^2 \Omega}{\delta \Delta^\mu_\alpha(\vv{q};\vv{k}_1) 
    \delta \Delta^\nu_\beta(\vv{q};\vv{k}_2)} \klsr{}_{\vv{q}=0} \nonumber \\
    &= V U^{-1}(\vv{k}_1,\vv{k}_2) \delta_{\alpha\beta} \delta_{\mu\nu} - \Pi_{\alpha\mu,\beta\nu}(\vv{k}_1) \delta_{\vv{k}_1,\vv{k}_2} \,,
\end{align}
where 
\begin{align}
    \Pi_{\alpha\mu,\beta\nu}(\vv{k}_1) =  \sum_{n,m=1}^{N_{\mathrm{B}}} \frac{\mathrm{Re}[(A_{\alpha,\mu}(\vv{k}_1))_{nm} (A_{\beta,\nu}(\vv{k}_1))_{nm}^\ast]}{E_{\vv{k}_1n} + E_{-\vv{k}_1m}} \,.
\end{align}
To find the inverse of this matrix with respect to the multi-indices $(\vv{k}_1,\alpha,\mu)$ and $(\vv{k}_2,\beta,\nu)$ we think of the full space as a tensor product of momentum space (index $\vv{k}$) and internal space (indices $\alpha,\mu$) and denote the corresponding basis vectors by $\ket{\vv{k}_1\alpha\mu}$ and $\ket{\vv{k}_2\beta\nu}$. Define the matrix $\hat{M}$ via
\begin{align}
    \frac{\delta^2 \Omega}{\delta \Delta^\mu_\alpha(\vv{q};\vv{k}_1) 
    \delta \Delta^\nu_\beta(\vv{q};\vv{k}_2)} \klsr{}_{\vv{q}=0} \eqqcolon \bra{\vv{k}_1\alpha\mu}\hat{M}\ket{\vv{k}_2\beta\nu} \,.
    \label{eq:definition-matrix-M}
\end{align}
and the matrices $\hat{U}$ and $\hat{\Pi}$ via
\begin{align}
    U(\vv{k}_1,\vv{k}_2) \delta_{\alpha\beta} \delta_{\mu\nu} &\eqqcolon \bra{\vv{k}_1\alpha\mu}\hat{U}\ket{\vv{k}_2\beta\nu} \\
    \Pi_{\alpha\mu,\beta\nu}(\vv{k}_1) \delta_{\vv{k}_1,\vv{k}_2} &\eqqcolon \bra{\vv{k}_1\alpha\mu}\hat{\Pi}\ket{\vv{k}_2\beta\nu} \,,
\end{align}
such that $\hat{M} = V\hat{U}^{-1} - \hat{\Pi}$. Consequently, the matrix representation of $\hat{U}$ is given by
\begin{align}
    \hat{U} = \sum_{\alpha,\mu} \sum_{\vv{k},\vv{k}'} U(\vv{k},\vv{k}') \ket{\vv{k}\alpha\mu} \bra{\vv{k}'\alpha\mu}\,,
\end{align}
and the matrix representation of $\hat{\Pi}$ is given by
\begin{align}
    \hat{\Pi} = \sum_{\alpha,\beta} \sum_{\mu,\nu} \sum_{\vv{k}} \Pi_{\alpha\mu,\beta\nu}(\vv{k}) \ket{\vv{k}\alpha\mu}\bra{\vv{k}\beta\nu} \,.
\end{align}
Hence, it is useful to express the inverse $\hat{M}^{-1}$ as
\begin{align}
    \hat{M}^{-1} = (V\hat{U}^{-1} - \hat{\Pi})^{-1} = \frac{1}{V}\klr{\hat{1} - \frac{1}{V} \hat{U}\hat{\Pi}}^{-1}\hat{U}  \,.
    \label{eq:inverse-m-factors}
\end{align}
\indent If we collect all the results, it turns out that we can express the functional superfluid weight as
\begin{align}                       D_{\mathrm{s},ij}^{\mathrm{func.}} =& \frac{1}{V} \sum_{\mu,\nu = \mathrm{R},\mathrm{I}} \sum_{\vv{k},\vv{k}'} \sum_{\alpha,\beta = 1}^{N_{\mathrm{B}}} \mathcal{S}_{\alpha i,\mu}(\vv{k}) M^{-1}_{\alpha\mu,\beta\nu}(\vv{k},\vv{k}') \mathcal{S}_{\beta j,\nu}(\vv{k}'), \label{eq:functional-superfluid-weight-wrt-skew-matrices}
\end{align}
where 
\begin{align} 
    \mathcal{S}_{\alpha i,\mu}(\vv{k}) = \sum_{n,m=1}^{N_{\mathrm{B}}} \frac{\mathrm{Re}[(A_{\alpha,\mu}(\vv{k}))_{nm} (B_i(\vv{k}))^\ast_{nm}]}{E_{\vv{k}n} + E_{-\vv{k}m}}
\end{align}
and 
\begin{align}
    M^{-1}_{\alpha\mu,\beta\nu}(\vv{k},\vv{k}') = \bra{\vv{k}\alpha\mu} \hat{M}^{-1}\ket{\vv{k}'\beta\nu} \,.
\end{align}
This formula allows a first observation: The numbers $\mathcal{S}_{\vv{k}\alpha,i}$ contain derivatives of the single-particle Hamiltonian. Therefore, these should contain quantum-geometrical information. We will examine this statement in more detail within the isolated narrow-bands limit in the next section.

Now we have two options how to proceed. 
The first option works if the interaction factorizes as 
\begin{align}
U(\vv{k},\vv{k}') = \sum_{i=1}^d u_{1,i}(\vv{k}) u_{2,i}(\vv{k}')\,.
\end{align}
Then one can interpret the interaction as a rank-$2 d  N_{\mathrm{B}}$ update for the inverse of the matrix $\hat{\Pi}$ and we can apply the Sherman-Morrison-Woodbury formula~\cite{woodbury1950inverting} (cf.\ Appendix~\ref{sec:woodbury}) to obtain an exact result without truncations. This is a convenient method we will use in Sec.~\ref{sec:kane-mele-d-wave} for the chiral $d$-wave superconducting state in a Kane-Mele-type model.

The second option consists of expanding the second factor of Eq.~\eqref{eq:inverse-m-factors} in a geometric series. Then $\hat{M}^{-1}$ is given by
\begin{align}
    \hat{M}^{-1} = \frac{1}{V}\hat{U} + \frac{1}{V^2}\hat{U} \hat{\Pi} \hat{U} + \frac{1}{V^3} \hat{U} \hat{\Pi} \hat{U} \hat\Pi \hat{U} + \hdots \,.
\end{align}
If the characteristic interaction strength is small, it should be sufficient to truncate the geometric series at some point as the role of higher-order terms is negligible. 
We then obtain a series expansion for the functional superfluid weight, reading as
\begin{widetext}
    \begin{align}   D_{\mathrm{s},ij}^{\mathrm{func.}} =& \frac{1}{V^2}\!\! \sum_{\mu = \mathrm{R,I}} \sum_{\vv{k},\vv{k}',\alpha} \mathcal{S}_{\alpha i,\mu}(\vv{k}) U(\vv{k},\vv{k}') \mathcal{S}_{\alpha j,\mu}(\vv{k}') \\
    &- \sum_{n=1}^{\infty} \frac{1}{V^{2+n}} \sum_{\substack{\mu_1,\hdots,\mu_n = \mathrm{R},\mathrm{I}\\ \vv{k},\vv{k}_1,\hdots,\vv{k}_n,\vv{k}'\\ \alpha_1,\hdots,\alpha_n}} \mathcal{S}_{\alpha_1 i,\mu_1}(\vv{k}) U(\vv{k},\vv{k}_1) \Pi_{\alpha_1\mu_1,\alpha_2\mu_2}(\vv{k}_1) U(\vv{k}_1,\vv{k}_2) \hdots \Pi_{\alpha_{n-1}\mu_{n-1},\alpha_n\mu_n}(\vv{k}_n) U(\vv{k}_n,\vv{k}')  \mathcal{S}_{\alpha_n j,\mu_n}(\vv{k}') \,.
     \nonumber
\end{align}
\end{widetext}
%

\section{Superfluid weight for the case of isolated narrow bands}\label{sec:isolated-band-limit}

In the last section we have derived formulas for the different contributions to the superfluid weight of multiband BCS theory for an arbitrary attractive effective two-particle interaction. 
It is clear that the conventional contribution is driven by the curvature of the bands [cf.\ Eq.~\eqref{eq:superfluid-weight-conv-def}]. 
The physical mechanism behind the other contributions is due to quantum geometry as these contributions contain derivatives of the single-particle eigenfunctions. 
This becomes evident within the isolated-bands limit. 

In the isolated-bands limit we assume that the bands of the Hamiltonian are well separated, i.e., the band gap between bands is larger than other energy scales~\footnote{It may be possible to reformulate this assumption, such that one assumes that some subset of $N_{\mathrm{B}'} \le N_{\mathrm{B}}$ bands is isolated with respect to the other bands since we expect the calculation to be very similar to the one presented in this section here. The key difference is that $\mathcal{D}_{n,\vv{k}}(\vv{q})$ defined in Eq.~\eqref{eq:definition-Dnkq} does not represent a complex number but rather a complex-valued ($N_{\mathrm{B}'} \times N_{\mathrm{B}'}$) matrix.}. 
Furthermore, we assume that the energy bands of the single-particle Hamiltonian are smooth in $\vv{k}$ and that the gap function is proportional to the identity matrix in band space. 
Then, let us take a look at the $n$th energy band of the Hamiltonian $H(\vv{k})$, denote its energy by $\varepsilon_{n}(\vv{k})$, and denote the corresponding (orthonormalized) single-particle eigenstate by $\ket{\psi_n(\vv{k})} \coloneqq [S(\vv{k})]_{\cdot,n}$
and 
\begin{align}
    \mathcal{D}_{n,\vv{k}}(\vv{q}) \coloneqq \bra{\psi_n(\vv{k}-\vv{q})} \Delta(\vv{k}) \ket{\psi_n(\vv{k}+\vv{q})} \,.
    \label{eq:definition-Dnkq}
\end{align}
Since we work within the isolated-bands limit, we have 
\begin{align}
    \bigoplus_{n=1}^{N_{\mathrm{B}}} \mathcal{H}_{n,\vv{k}}(\vv{q}) \approx \mathcal{H}_{\vv{k}}(\vv{q})\,,
\end{align}
where the BdG Hamiltonian associated to the $n$th band reads as
\begin{align}
    \mathcal{H}_{n,\vv{k}}(\vv{q}) = \matr{\varepsilon_{n}(\vv{k} - \vv{q}) - \mu & \mathcal{D}_{n,\vv{k}}(\vv{q})\\ \mathcal{D}^\dagger_{n,\vv{k}}(\vv{q}) & -\varepsilon_{n}(\vv{k} + \vv{q}) + \mu} \,.
\end{align}
Our goal now consists of finding the quasiparticle eigenvalues of the single-particle diagonalized BdG Hamiltonian $\mathcal{H}_{\vv{k}}(\vv{q})$ within this limit. First, we observe that the BdG Hamiltonian squared is given by
\begin{widetext}
\begin{align}
    \mathcal{H}_{n,\vv{k}}^2(\vv{q}) &= \matr{(\varepsilon_{n}(\vv{k} - \vv{q}) - \mu)^2 + \mathcal{D}_{n,\vv{k}}(\vv{q}) \mathcal{D}^\dagger_{n,\vv{k}}(\vv{q}) & (\varepsilon_n(\vv{k}-\vv{q}) - \varepsilon_n(\vv{k}+\vv{q})) \mathcal{D}_{n,\vv{k}}(\vv{q})\\ (\varepsilon_n(\vv{k}-\vv{q}) - \varepsilon_n(\vv{k}+\vv{q})) \mathcal{D}^\dagger_{n,\vv{k}}(\vv{q}) & (\varepsilon_{n}(\vv{k} + \vv{q}) - \mu)^2 + \mathcal{D}_{n,\vv{k}}(\vv{q}) \mathcal{D}^\dagger_{n,\vv{k}}(\vv{q})} \,.
\end{align}
We denote the diagonal part by
\begin{align}
    \tilde{\mathcal{H}}_{n,\vv{k}}^2(\vv{q}) \coloneqq \matr{(\varepsilon_{n}(\vv{k} - \vv{q}) - \mu)^2 + \mathcal{D}_{n,\vv{k}}(\vv{q}) \mathcal{D}^\dagger_{n,\vv{k}}(\vv{q}) & 0\\ 0 & (\varepsilon_{n}(\vv{k} + \vv{q}) - \mu)^2 + \mathcal{D}_{n,\vv{k}}(\vv{q}) \mathcal{D}^\dagger_{n,\vv{k}}(\vv{q})}
\end{align}
and the off-diagonal part by
\begin{align}
    \Lambda_{n,\vv{k}}(\vv{q}) &= \matr{0 & (\varepsilon_n(\vv{k}-\vv{q}) - \varepsilon_n(\vv{k}+\vv{q})) \mathcal{D}_{n,\vv{k}}(\vv{q})\\ (\varepsilon_n(\vv{k}-\vv{q}) - \varepsilon_n(\vv{k}+\vv{q})) \mathcal{D}^\dagger_{n,\vv{k}}(\vv{q}) & 0} \,.
\end{align}
\end{widetext}
Performing a Taylor expansion due to the assumed smoothness of the bands we find that the diagonal part dominates for small $\vv{q}$ here as for $\vv{q} \to 0$ the off-diagonal elements are of order
\begin{align}
    &\klr{\varepsilon_n(\vv{k}-\vv{q}) - \varepsilon_n(\vv{k}+\vv{q})}\mathcal{D}^{(\dagger)}_{n,\vv{k}}(\vv{q}) \nonumber \\
    &\qquad\qquad = -2\Delta_n^{(\ast)}(\vv{k}) \klr{\nabla_{\vv{k}} \varepsilon_n(\vv{k}) \cdot \vv{q}} \,.
\end{align}
This allows us to treat $\Lambda_{n,\vv{k}}(\vv{q})$ as a perturbation, i.e., by using perturbation theory, we can analyze the corrections to the eigenvalues systematically. The eigenvalues of $\tilde{\mathcal{H}}_{n,\vv{k}}^2(\vv{q})$ can be easily calculated and are given by
\begin{align}
    \big(E_{n,\vv{k}}^{(0)}(\vv{q})\big)^2 = (\varepsilon_{n}(\vv{k} \pm \vv{q}) - \mu)^2 + \mathcal{D}_{n,\vv{k}}(\vv{q}) \mathcal{D}^\dagger_{n,\vv{k}}(\vv{q}) \,.
\end{align}
Therefore, the grand potential \eqref{eq:grand-potential-mean-field-bcs} has the following form at zeroth-order perturbation theory:
\begin{align}
    &\Omega^{(0)}(\vv{q})\!=\! \frac{1}{2}\! \sum_{\vv{k}} \!\tr(\varepsilon_{\vv{k}-\vv{q}}\!-\!\mu) 
    \!+\! \frac{V}{2} \!\sum_{\vv{k},\vv{k}'} U^{-1}(\vv{k},\vv{k}') \bar{\Delta}_{\alpha}(\vv{k}) \Delta_{\alpha}(\vv{k}')  \nonumber \\
    &\quad -\frac{1}{4} \sum_{n,\vv{k}} \Big(\sqrt{(\varepsilon_{n}(\vv{k} + \vv{q}) - \mu)^2 + \mathcal{D}_{n,\vv{k}}(\vv{q}) \mathcal{D}^\dagger_{n,\vv{k}}(\vv{q})} \nonumber \\
    &\quad\quad\quad\quad + \sqrt{(\varepsilon_{n}(\vv{k} - \vv{q}) - \mu)^2 + \mathcal{D}_{n,\vv{k}}(\vv{q}) \mathcal{D}^\dagger_{n,\vv{k}}(\vv{q})}\Big) \,.
\end{align}
Since $\Delta(\vv{k})$ is diagonal, we obtain
\begin{align}
    \mathcal{D}_{n,\vv{k}}(0) \mathcal{D}^\dagger_{n,\vv{k}}(0) = |\Delta(\vv{k})|^2
\end{align}
and
\begin{align}
    \partial_{q_i} \left.\klr{\mathcal{D}_{n,\vv{k}}(\vv{q}) \mathcal{D}^\dagger_{n,\vv{k}}(\vv{q})} \right|_{\vv{q}=0} = 0 
    \label{eq:first-derivative-vanishes-D}\,.
\end{align}
%

\subsection{Conventional and geometrical contribution}

For the following, let us denote by
\begin{align}
    E^{(0)}_{n,\vv{k}} = \sqrt{(\varepsilon_n(\vv{k}) - \mu)^2 + |\Delta(\vv{k})|^2} \,.
\end{align}
Since the conventional superfluid weight is proportional to the effective mass, at zero-order perturbation theory it can be identified as
\begin{align}
    D_{\mathrm{s},ij}^{\mathrm{conv},(0)} = \frac{1}{2V} \sum_{n,\vv{k}} \frac{\partial^2 \varepsilon_n(\vv{k})}{\partial{k_i} \partial{k_j}} \klr{1 - \frac{\varepsilon_n(\vv{k}) - \mu}{E^{(0)}_{n,\vv{k}}}} \,. 
\end{align}
The remaining terms account for the geometrical contribution to the superfluid weight. By applying Eq.~\eqref{eq:first-derivative-vanishes-D}, this contribution at zeroth-order perturbation theory is given by
\begin{align}
    D_{\mathrm{s},ij}^{\mathrm{geom,(0)}} =& -\frac{1}{4V} \sum_{n,\vv{k}} \frac{1}{E^{(0)}_{n,\vv{k}}} \Bigg[ \frac{\partial^2 \big(\mathcal{D}_{n,\vv{k}}(\vv{q}) \mathcal{D}^\dagger_{n,\vv{k}}(\vv{q})\big)\big|_{\vv{q}=0}}{\partial{q_i} \partial{q_j}} \nonumber \\
    & + \frac{\partial\varepsilon_n(\vv{k})}{\partial k_i} \frac{\partial \varepsilon_n(\vv{k})}{\partial k_j} \klr{1 +  \frac{|\Delta(\vv{k})|^2}{\big(E^{(0)}_{n,\vv{k}}\big)^2}}  \Bigg] \,.
\end{align}
Now, by calculating the second derivative of $\braket{\psi_n} = 1$ on both sides, one can find the following relation:
\begin{align}
    &\braket{\partial_{k_i} \partial_{k_j} \psi_n(\vv{k})}{\psi_n(\vv{k})} + \braket{\psi_n(\vv{k})}{\partial_{k_i} \partial_{k_j} \psi_n(\vv{k})} \nonumber \\[8pt]
    &= -\braket{\partial_{k_i} \psi_n(\vv{k})}{\partial_{k_j} \psi_n(\vv{k})} - \braket{\partial_{k_j} \psi_n(\vv{k})}{\partial_{k_i} \psi_n(\vv{k})} \,.
\end{align}
Using this, the second derivative of $\mathcal{D}_{n,\vv{k}} \mathcal{D}_{n,\vv{k}}^\dagger$ at $\vv{q}=0$ reads as
\begin{align}
    &\frac{\partial^2 \big(\mathcal{D}_{n,\vv{k}}(\vv{q}) \mathcal{D}^\dagger_{n,\vv{k}}(\vv{q})\big)}{\partial{q_i} \partial{q_j}}\klsr{}_{\vv{q}=0} = -4|\Delta(\vv{k})|^2 \big(\braket{\partial_{k_i}\psi_n}{\partial_{k_j}\psi_n} \nonumber \\[8pt]
    &\quad + \braket{\partial_{k_i}\psi_n}{\psi_n} \braket{\psi_n}{\partial_{k_j}\psi_n} + (i \leftrightarrow j) \big) \,.
\end{align}
This corresponds to the quantum metric of the $n$th band
\begin{align}
    g_{ij}^{(n)}\!=\! \frac{1}{2}\kle{\braket{\partial_i\psi_n}{\partial_j\psi_n}- \braket{\partial_i\psi_n}{\psi_n} \braket{\psi_n}{\partial_j\psi_n} + \klr{i \leftrightarrow j}}\,,
    \label{eq:quantum-metric-def}
\end{align}
i.e., the geometrical contribution becomes
\begin{align}
    D_{\mathrm{s},ij}^{\mathrm{geom},(0)} =& \frac{1}{V} \sum_{n,\vv{k}} \frac{1}{E^{(0)}_{n,\vv{k}}} \Bigg[ |\Delta(\vv{k})|^2 g^{(n)}_{ij}(\vv{k}) \nonumber \\
    &- \frac{1}{4} \frac{\partial\varepsilon_n(\vv{k})}{\partial k_i} \frac{\partial \varepsilon_n(\vv{k})}{\partial k_j} \klr{1 +  \frac{|\Delta(\vv{k})|^2}{\big(E^{(0)}_{n,\vv{k}}\big)^2}}  \Bigg]\,,\label{eq:explicitgeom}
\end{align}
This shows that the quantum metric plays an important role in the superfluid weight of systems with narrow bands. Additionally, it also depends on the group velocity.

\subsection{Functional contribution}

Analogously, we can obtain an expression for the functional superfluid weight within the isolated-bands limit. In the last section, it has been already mentioned that this contribution to the superfluid weight contains derivatives of the single-particle Hamiltonian, i.e., derivatives of the eigenfunctions of the single-particle Hamiltonian. Therefore, we expect to observe the occurrence of quantum geometry. To make this statement more precise we calculate the superfluid weight within the isolated-bands limit and perform a similar calculation as it has been done for the conventional and geometrical superfluid weight earlier in this section. Note that it is important to employ the assumption that the gap function is proportional to the identity matrix after calculating all the necessary derivatives.

According to Eq.~\eqref{eq:superfluid-weight-bcs-trs} we need the derivative of the grand potential with respect to $q_i$ which is given by 
\begin{align}
    \frac{\partial \Omega^{(0)}}{\partial q_i} &\supset -\frac{1}{4} \sum_{n,\vv{k}} \Bigg(\frac{\partial_{q_i} \klr{\mathcal{D}_{n,\vv{k}}(\vv{q})\mathcal{D}^\dagger_{n,\vv{k}}(\vv{q})}}{\sqrt{\klr{\varepsilon_n(\vv{k}) - \mu}^2 + \mathcal{D}_{n,\vv{k}}(\vv{q})\mathcal{D}^\dagger_{n,\vv{k}}(\vv{q})}} \nonumber \\[8pt]
    &+ \frac{\partial \varepsilon_n(\vv{k})}{\partial k_i} \frac{2(\varepsilon_n(\vv{k}) - \mu)}{\sqrt{(\varepsilon_n(\vv{k})  - \mu)^2 + \mathcal{D}_{n,\vv{k}}(\vv{q}) \mathcal{D}^\dagger_{n,\vv{k}}(\vv{q})}} \Bigg)\,,
\end{align}
whereby \enquote{$\supset$} indicates dropped terms that do not contribute when calculating the functional derivative with respect to the gap function. 
Denote by $\klg{\ket{\alpha}}$ the canonical basis of the Hilbert space spanned by the eigenfunctions of the single-particle Hamiltonian and define 
\begin{align}
    R^{(1)}_{n,\alpha,i}(\vv{k}) &\coloneqq \braket{\psi_n}{\alpha}\braket{\alpha}{\partial_{k_i}\psi_n} \qquad \label{eq:definition-R-1}
\end{align}
and
\begin{align}
    R^{(2)}_{n,\alpha,i}(\vv{k}) &\coloneqq \bra{\partial_{k_i} \psi_n}\ket{\psi_n} \braket{\psi_n}{\alpha} \braket{\alpha}{\psi_n} \,.\label{eq:definition-R-2}
\end{align}
After some calculation it turns out that we can express the necessary derivatives as 
\begin{align}
    &\frac{\delta}{\delta \Delta^{\mathrm{R,I}}_\alpha(\vv{q};\vv{k}')} \klr{\partial_{q_i} \klr{\mathcal{D}_{n,\vv{k}}(\vv{q})\mathcal{D}^\dagger_{n,\vv{k}}(\vv{q})}} \klsr{}_{\vv{q}=0}\nonumber\\[8pt] &\quad= -4\Delta^{\mathrm{I},\mathrm{R}}(\vv{k}) \delta_{\vv{k},\vv{k}'} \mathrm{Im}(R^{(1)}_{n,\alpha,i}(\vv{k}) + R^{(2)}_{n,\alpha,i}(\vv{k})) \,,
\end{align}
\begin{align}
    \frac{\delta \klr{\mathcal{D}_{n,\vv{k}}(\vv{q})\mathcal{D}^\dagger_{n,\vv{k}}(\vv{q})}}{\delta \Delta_\alpha^{\mathrm{R,I}}(\vv{q};\vv{k}')}  \klsr{}_{\vv{q}=0} = \pm 2 \Delta^{\mathrm{R,I}}(\vv{k}) \delta_{\vv{k},\vv{k}'} \kls{\braket{\psi_n(\vv{k})}{\alpha}}^2 \,.
\end{align}
Therefore, the zeroth order of perturbation theory dictates a functional contribution to the superfluid weight of
\begin{widetext}
\begin{align}
    D_{\mathrm{s},ij}^{\mathrm{func},(0)} &= \frac{1}{V} \!\!\!\!\sum_{n,n',\vv{k},\vv{k}',\alpha,\beta,\mu,\nu}\! \Bigg[\!\frac{M^{-1}_{\alpha\mu,\beta\nu}(\vv{k},\vv{k}') \Delta^{\bar{\mu}}(\vv{k}) \Delta^{\bar{\nu}}(\vv{k}') \mathrm{Im}(R^{(1)}_{n,\alpha,i}(\vv{k}) \!+\! R^{(2)}_{n,\alpha,i}(\vv{k})) \mathrm{Im}(R^{(1)}_{n',\beta,j}(\vv{k}') \!+\! R^{(2)}_{n',\beta,j}(\vv{k}'))}{\sqrt{(\varepsilon_n(\vv{k}) - \mu)^2 + |\Delta(\vv{k})|^2} \sqrt{(\varepsilon_{n'}(\vv{k}') - \mu)^2 + |\Delta(\vv{k}')|^2}} \nonumber \\
    &\hspace{-0.4cm}+ \frac{M^{-1}_{\alpha\mu,\beta\nu}(\vv{k},\vv{k}') \Delta^\mu(\vv{k}) \Delta^\nu(\vv{k}') \klr{\varepsilon_n(\vv{k}) - \mu} \klr{\varepsilon_{n'}(\vv{k}') - \mu} \kls{\braket{\psi_n(\vv{k})}{\alpha}}^2 \kls{\braket{\psi_{n'}(\vv{k}')}{\beta}}^2 }{\klr{(\varepsilon_n(\vv{k}) - \mu)^2 + |\Delta(\vv{k})|^2}^{3/2} \klr{(\varepsilon_{n'}(\vv{k}') - \mu)^2 + |\Delta(\vv{k}')|^2}^{3/2}} \frac{\partial \varepsilon_n(\vv{k})}{\partial k_i} \frac{\partial \varepsilon_{n'}(\vv{k}')}{\partial k_j'} \Bigg]  ,\label{eq:explicitfunc}
\end{align}   
\end{widetext}
where $\bar{\mu} = \mathrm{R}$ if $\mu = \mathrm{I}$ and $\bar{\mu} = \mathrm{I}$ if $\mu = \mathrm{R}$. The first term is of geometrical nature as it depends on derivatives of the single-particle eigenfunctions. 
The second term depends on derivatives of the single-particle eigenvalues only and represents a non-geometrical contribution. 

\subsection{Functional contribution and Wilczek-Zee connection}

Let us take a closer look at the first term and expand the standard basis $\klg{\ket{\alpha}}$ with respect to the eigenbasis $\klg{\ket{\psi_m(\vv{k})}}$ of the single-particle Hamiltonian
\begin{align}
    \ket{\alpha} = \sum_{m=1}^{N_{\mathrm{B}}} \phi_m^\alpha(\vv{k}) \ket{\psi_m(\vv{k})} \,.
    \label{eq:expansion-coeffs-eigenbasis}
\end{align}
Consequently, the sum $R^{(1)}_{n,\alpha,i}(\vv{k}) + R^{(2)}_{n,\alpha,i}(\vv{k})$ is given by
\begin{align}
    &R^{(1)}_{n,\alpha,i}(\vv{k}) + R^{(2)}_{n,\alpha,i}(\vv{k}) = \sum_{m \neq n} \phi_n^\alpha \bar{\phi}_m^\alpha \braket{\psi_m}{\partial_{k_i}\psi_n}\,.
\end{align}
\\

We identify the Wilczek-Zee connection $e_{i,m}^{(n)}$ of the $n$th band defined by \cite{wilczek1984appearance}
\begin{align}
    e_{i,m}^{(n)} \coloneqq i\braket{\psi_m}{\partial_{k_i}\psi_n}\,.
    \label{eq:wilczek-zee-connection-definition}
\end{align}
It measures essentially the ability of the $n$th state to change into the $m$th state \cite{romero2024n}. 
It is convenient to express this quantity in terms of projectors (cf.\ Ref.~\cite{Bouhon2023,mitscherling2024gauge}) as 
\begin{align}
    e^{(n)}_{i,m} = i \tr((\partial_{k_i} \hat{P}_{nm}) \hat{P}_m)\,,
\end{align}
where
\begin{align}
    \hat{P}_{nm} = \ket{\psi_n} \bra{\psi_m}\,, \quad 
    \hat{P}_m = \ket{\psi_m} \bra{\psi_m} \,.
    \label{eq:projector-definition}
\end{align}
Due to $\mathrm{Im}(a)\,\mathrm{Im}(b) = \frac{1}{2}\mathrm{Re}\klr{a\bar{b} - ab}$, we can express the functional contribution within the isolated-bands limit as
\begin{widetext}
\begin{align}
    D_{\mathrm{s},ij}^{\mathrm{func},(0)} =& \frac{1}{V} \sum_{\substack{n,n',\vv{k},\vv{k}'\\ \alpha,\beta}}  F_{nn',\alpha\beta}(\vv{k},\vv{k}') \Bigg[\,
    \frac{1}{2}\,\sum_{m\neq n,m' \neq n'}\mathrm{Re}\klr{O^{(1)}_{nn'mm',\alpha\beta}(\vv{k},\vv{k}')  e_{i,m}^{(n)}(\vv{k}) \bar{e}_{j,m'}^{(n')}(\vv{k}') + O^{(2)}_{nn'mm',\alpha\beta}(\vv{k},\vv{k}') e_{i,m}^{(n)}(\vv{k}) \bar{e}_{j,n'}^{(m')}(\vv{k}')} \nonumber \\
    &+ \frac{(\varepsilon_n(\vv{k}) - \mu) (\varepsilon_{n'}(\vv{k}') - \mu)  \kls{\braket{\psi_n(\vv{k})}{\alpha}}^2 \kls{\braket{\psi_{n'}(\vv{k}')}{\beta}}^2}{(E_{n,\vv{k}}^{(0)})^2 (E_{n',\vv{k}'}^{(0)})^2} \frac{\partial \varepsilon_n(\vv{k})}{\partial k_i} \frac{\partial \varepsilon_{n'}(\vv{k}')}{\partial k_j'} \Bigg] \,,
    \label{eq:exact-functional-contribution-isolated-band}
\end{align}
\end{widetext}
where 
\begin{align}
    O^{(1)}_{nn'mm',\alpha\beta}(\vv{k},\vv{k}') &\coloneqq  \phi_n^\alpha(\vv{k}) \bar{\phi}_m^{\alpha}(\vv{k}) \bar{\phi}_{n'}^{\beta}(\vv{k}') \phi_{m'}^{\beta}(\vv{k}')\,,\\
    O^{(2)}_{nn'mm',\alpha\beta}(\vv{k},\vv{k}') &\coloneqq  \phi_n^\alpha(\vv{k}) \bar{\phi}_m^{\alpha}(\vv{k}) \phi_{n'}^{\beta}(\vv{k}') \bar{\phi}_{m'}^{\beta}(\vv{k}')
\end{align}
and
\begin{align}
   F_{nn', \alpha\beta}(\vv{k},\vv{k}') \coloneqq \frac{M^{-1}_{\alpha\mu,\beta\nu}(\vv{k},\vv{k}') \Delta^\mu(\vv{k}) \Delta^\nu(\vv{k}')}{E^{(0)}_{n,\vv{k}} E^{(0)}_{n',\vv{k}'}} \,.
\end{align}
Note that the products of the Wilczek-Zee connections and the expansion coefficients in Eq.~\eqref{eq:exact-functional-contribution-isolated-band} are gauge invariant, i.e., the expression of the functional superfluid weight here is also gauge invariant. 

\subsection{Functional contribution and two-point fidelity magnitude}

Let us analyze the functional contribution in more detail and assume that $D_{\mathrm{s},ij}^{\mathrm{func},(0)} > 0$ is positive, i.e., we assume that it lowers the total superfluid weight \eqref{eq:superfluid-weight-bcs-trs}. This is, for example, always the case for the diagonal elements $D_{\mathrm{s},ii}^{\mathrm{func},(0)}$ of the functional superfluid weight since, by solving the gap equation, we work at a minimum of the grand potential. For nondiagonal elements $i \neq j$ this is not always guaranteed. \\
\indent Hence, if we find an upper bound for the functional contribution, we have a lower bound for the total contribution. Let
\begin{align}
    F_{nn'}(\vv{k},\vv{k}') = \sum_{\alpha,\beta=1}^{N_{\mathrm{B}}} \frac{M^{-1}_{\alpha\mu,\beta\nu}(\vv{k},\vv{k}') \Delta^\mu(\vv{k}) \Delta^\nu(\vv{k}')}{E^{(0)}_{n,\vv{k}} E^{(0)}_{n',\vv{k}'}}\,.
\end{align}
Using the triangle inequality, we can estimate the functional superfluid weight as
\begin{align}
    D_{\mathrm{s},ij}^{\mathrm{func},(0)} &\le \frac{1}{V} \sum_{n,n',\vv{k},\vv{k}'} |F_{nn'}(\vv{k},\vv{k}')| \Bigg[ \sum_{\substack{m\neq n,\\m' \neq n'}} \kls{e_{i,m}^{(n)}(\vv{k}) \bar{e}_{j,m'}^{(n')}(\vv{k}')} \nonumber \\
    &\hspace{-0.4cm}+ \frac{\kls{\varepsilon_n(\vv{k}) - \mu} \kls{\varepsilon_{n'}(\vv{k}') - \mu}}{(E_{n,\vv{k}}^{(0)})^2 (E_{n',\vv{k}'}^{(0)})^2} \kls{\frac{\partial \varepsilon_n(\vv{k})}{\partial k_i} \frac{\partial \varepsilon_{n'}(\vv{k}')}{\partial k_j'}} \Bigg]\,,
    \label{eq:bound-func-temp}
\end{align}    
where we have used the fact that the expansion coefficients satisfy $|\phi_n^\alpha| \le 1$. We identify the two-point fidelity magnitude
\begin{align}
    \zeta_{ij}^{(nn')}(\vv{k},\vv{k}') \coloneqq \sum_{m\neq n,m' \neq n'} \kls{e_{i,m}^{(n)}(\vv{k}) e_{j,m'}^{(n')}(\vv{k}')}
    \label{eq:two-point-fidelity-tensor-definition} 
\end{align}
and obtain the upper bound 
\begin{align}
    D_{\mathrm{s},ij}^{\mathrm{func},(0)} &\le \frac{1}{V} \!\!\!\!\sum_{n,n',\vv{k},\vv{k}'}\!\!\!\! |F_{nn'}(\vv{k},\vv{k}')| \kle{ \zeta_{ij}^{(nn')}(\vv{k},\vv{k}') + C_{ij}(\vv{k},\vv{k}') } \,,
    \label{eq:estimation-func-superfluid-weight-fidelity}
\end{align}
where $C_{ij}(\vv{k},\vv{k}')$ represents the second line of Eq.~\eqref{eq:bound-func-temp} which does not depend on quantum geometry and is equal to zero if a flat band is present. Therefore, we can see that the functional contribution is driven by the presence of a nonzero two-point fidelity magnitude. However, keep in mind that this represents an upper bound only which may also result in a negative lower bound for the total superfluid weight in some cases. 

The first-order correction to the quasiparticle eigenvalues (and, therefore, to the superfluid weight) is zero, i.e., until now, we have considered perturbation theory of the BdG Hamiltonian up to the first-order correction. In principle, it is also possible to include higher-order corrections by taking the matrix $\Lambda_{n,\vv{k}}(\vv{q})$ into consideration. However, if the bands are flat, we have $\Lambda_{n,\vv{k}}(\vv{q}) = 0$, i.e., the results presented in this section are exact (within the isolated-bands limit). Nevertheless, as long as our bands are not flat, since the matrix $\Lambda_{n,\vv{k}}(\vv{q})$ contains the single-particle eigenfunctions, we expect to obtain quantum geometrical corrections for higher orders of perturbation theory and we assume that this approach reproduces and further generalizes the results obtained by Jiang and Barlas in Ref.~\cite{jiang2024geometric}. \\

\subsection{Conventional flat-band limit and minimal quantum metric}\label{sec:isolated-flat-band-limit}

The goal of this section consists of highlighting the key differences between conventional and unconventional pairing scenarios and emphasizing to what extent the two-point fidelity magnitude may allow a quantum-geometrical interpretation of the appearance of the minimal quantum metric in the superfluid weight elaborated in Ref.~\cite{huhtinen2022revisiting}. 
In the following, we assume that the $n$th band of the system is flat and that the band gap between the flat band and other bands is larger than the band width, the interaction strength~$U_0$, and the order parameters.

According to the previous subsections in Sec.~\ref{sec:isolated-band-limit} and Appendix~\ref{sec:woodbury}, the superfluid weight within the isolated-bands limit of a uniform conventional $s$-wave order parameter can be approximated by 
\begin{widetext}
\begin{align}
    D_{\mathrm{s},ij} &\approx \frac{1}{V} \frac{|\Delta|^2}{E^{(0)}_n} \sum_{\vv{k}} g^{(n)}_{ij}(\vv{k}) - \underbrace{ \frac{|\Delta|^2}{2V\big( E^{(0)}_n \big)^2}  \!\!\!\sum_{\substack{\vv{k},\vv{k}',\alpha,\beta\\ m,m'\neq n}}\!\!\!\!   M^{-1}_{\alpha\beta} \, \mathrm{Re}\klr{O^{(1)}_{nnmm',\alpha\beta}(\vv{k},\vv{k}') e_{i,m}^{(n)}(\vv{k}) \bar{e}_{j,m'}^{(n)}(\vv{k}') \!+\! O^{(2)}_{nnmm',\alpha\beta}(\vv{k},\vv{k}')e_{i,m}^{(n)}(\vv{k}) \bar{e}_{j,n}^{(m')}(\vv{k}')}}_{\text{\enquote{minimal quantum metric} correction}} \,,
    \label{eq:superfluid-weight-special-case-isolated-flat-band}
\end{align}   
\end{widetext}
where $M^{-1}_{\alpha\beta}$ is defined in Eq.~\eqref{eq:conventional-M} and is independent of $\vv{k},\vv{k}'$. The effective single-particle Hamiltonian of a Cooper pair confined to the flat band is given by \cite{herzog2022many}
\begin{align}
    h_{\alpha\beta}(\vv{q}) = \sum_{\vv{k}} [\hat{P}_n(\vv{k}+\vv{q})]_{\alpha\beta} [\hat{P}_n(\vv{k})]_{\beta\alpha}\,,
\end{align}
where $\hat{P}_n$ denotes the projector into the $n$th band defined in Eq.~\eqref{eq:projector-definition} and $\alpha,\beta = 1,\hdots,N_{\mathrm{B}}$ denote the orbital indices. Let us assume that all orbitals are fixed at high-symmetry positions. Following the calculations of Ref.~\cite{huhtinen2022revisiting}, it turns out that 
\begin{align}
    \sum_{\beta = 1}^{N_{\mathrm{B}}} \partial_{k_i} h_{\alpha\beta}(0) = 0
\end{align}
needs to hold for all $i = 1,\hdots,\mathrm{dim}(\mathrm{BZ})$ and $\alpha = 1,\hdots,N_{\mathrm{B}}$. When inserting the definition of the projector $\hat{P}_n$ into the equation above, we find this condition to be equivalent to
\begin{align}
    \sum_{\vv{k}} \klr{\braket{\alpha}{\partial_{k_i} \psi_n} \braket{\psi_n}{\alpha} + \braket{\alpha}{\psi_n} \braket{\psi_n}{\alpha} \braket{\partial_{k_i}\psi_n}{\psi_n}} = 0 \,.\notag
\end{align}
We observe that this corresponds exactly to the sum of the quantities $R_{n,\alpha,i}^{(1)}$ and $R_{n,\alpha,i}^{(2)}$ defined in Eqs.~\eqref{eq:definition-R-1} and \eqref{eq:definition-R-2}.
Therefore, if we assume that we have an isolated flat band with orbitals fixed to high-symmetry positions and a conventional superconducting pairing mechanism, the functional contribution in Eq.~\eqref{eq:superfluid-weight-special-case-isolated-flat-band} is zero. 
Then, as expected, the superfluid weight is determined by the geometrical superfluid weight
\begin{align}
    D_{\mathrm{s},ij} \approx \frac{1}{V} \frac{|\Delta|^2}{E_n^{(0)}} \sum_{\vv{k}} g_{ij}^{\mathrm{min},(n)}(\vv{k}) \,,
\end{align}
i.e., in this limit, we reproduce~\cite{huhtinen2022revisiting} in the sense that the superfluid weight is solely determined by the minimal quantum metric. 
The advantage of our formula is its independence of the chosen basis for the orbitals. 
Furthermore, when we have unconventional pairing, the function we need when performing the $\vv{k}, \vv{k}'$ sums in the functional superfluid weight changes and additionally depends on the $\vv{k}$-dependent pairing functions $\Delta(\vv{k})$ and $\vv{k}$- and $\vv{k}'$-dependent pairing potential $U(\vv{k},\vv{k}')$. 
As the numerical calculations below show, the functional superfluid weight does not vanish in this case even though the orbitals are fixed to high-symmetry positions. 
Hence, in general, we have a non-zero functional contribution even if we choose a basis which corresponds to the minimal quantum metric.

Moreover, we can relate the functional contribution to multistate quantum geometry. As the first factor in Eq.~\eqref{eq:superfluid-weight-special-case-isolated-flat-band} is positive, we can find the following lower bound for the total superfluid weight by employing the estimation provided in Eq.~\eqref{eq:estimation-func-superfluid-weight-fidelity} for the functional superfluid weight 
\begin{align}
    D_{\mathrm{s},ij} &\gtrsim \frac{1}{V} \frac{|\Delta|^2}{E^{(0)}_n} \sum_{\vv{k}} g^{(n)}_{ij}(\vv{k}) - \!\frac{N_{\mathrm{B}}}{V^2} \frac{ U_0|\tilde{M}| |\Delta|^2}{\big( E^{(0)}_n \big)^2} \sum_{\vv{k},\vv{k}'}\! \zeta_{ij}^{(nn)}(\vv{k},\vv{k}') ,
    \label{eq:lower-bound-superfluid-weight-isolated-flat-band-conv}
\end{align}
where $\tilde{M} = \sum_{\alpha\beta} \big[(\check{\mathbbm{1}} + \check{\Sigma})^{-1}\big]_{\alpha\beta}$ [see also Eq.~\eqref{eq:conventional-M}]. This result implies that as soon as the order parameter is non-zero and we work in a basis in which the quantum metric does not become minimal, the superfluid weight is not solely determined by the quantum metric but rather experiences a correction due to the functional superfluid weight. 
This correction is maximally determined by the integrated two-point fidelity magnitude~\eqref{eq:two-point-fidelity-tensor-definition}, i.e., by the \enquote{similarity} of the eigenstates. 
%

\section{Application to a Kane-Mele-type model}\label{sec:application-km-model}

\subsection{Extended Kane-Mele Hamiltonian}

An interesting model that allows the study of geometrical nontrivial flat bands is an extended version of the Kane-Mele model with additional hoppings between the third- and fourth- nearest neighbors on a honeycomb lattice \cite{lau2022universal}:
\begin{align}
    &H_0 = t \!\sum_{\sigma, \kla{i,j}_1}\! c_{j\sigma}^\dagger c_{i\sigma} + t_2 \!\sum_{\sigma,\kla{i,j}_2}\! e^{i\sigma \varphi_{ij}} c_{j\sigma}^\dagger c_{i\sigma} \nonumber \\
    & + t_3 \!\sum_{\sigma,\kla{i,j}_3}\! c_{j\sigma}^\dagger c_{i\sigma} + t_4 \!\sum_{\sigma,\kla{i,j}_4}\! c_{j\sigma}^\dagger c_{i\sigma} + \sum_{\sigma, i} (-1)^i M c_{i\sigma}^\dagger c_{i\sigma} \,.
    \label{eq:extended-kane-mele-model}
\end{align}
The spin index is denoted by $\sigma = \pm 1 = \uparrow\downarrow$, $\kla{i,j}_n$ indicates summation over pairs of $n$th neighbors, $M$ is a staggered onsite potential, and $\varphi_{ij} = \pm\varphi$ is a hopping phase whose sign is chosen such that the Hamiltonian is time-reversal symmetric and therefore depends on the hopping direction and on the spin.
Figure~\ref{fig:extended-kane-mele-model-drawing} depicts the choice of the sign. In the limit $t_3 = t_4 = 0$ and $\varphi = \pi/2$ the Hamiltonian reduces to the model introduced by Kane and Mele in Refs.\ \cite{kane2005quantum,kane2005z}. 
\begin{figure}[t!]
    \centering
    \subfigure[]{\resizebox{0.99\linewidth}{!}{\begin{tikzpicture}
    
    \tikzset{
        atomA/.style={circle, draw=black, fill=black, inner sep=0pt, minimum size=6pt},
        atomB/.style={circle, draw=black, fill=white, inner sep=0pt, minimum size=6pt},
        vector/.style={->, red, thick, >=stealth},
        labelStyle/.style={red, thick}
    }
    
    \coordinate (A) at (0,0);
    \coordinate (B1) at (0.5,0.866);
    \coordinate (B2) at (0.5,-0.866);
    \coordinate (B3) at (-1,0);
    \coordinate (C1) at (1.5,0.866);
    \coordinate (C2) at (0,1.732);
    \coordinate (C3) at (-1.5,0.866);
    \coordinate (C4) at (-1.5,-0.866);
    \coordinate (C5) at (0,-1.732);
    \coordinate (C6) at (1.5,-0.866);
    \coordinate (D1) at (2,0);
    \coordinate (D2) at (-1,1.732);
    \coordinate (D3) at (-1,-1.732);
    \coordinate (E1) at (0.5,2.598);
    \coordinate (E2) at (2,1.732);
    \coordinate (E3) at (2,-1.732);
    \coordinate (E4) at (0.5,-2.598);
    \coordinate (E5) at (-2.5,-0.866);
    \coordinate (E6) at (-2.5,0.866);

    \coordinate (H1) at (3,0);
    \coordinate (H2) at (3,1.732);
    \coordinate (H3) at (3,-1.732);
    \coordinate (H4) at (3.5,0.866);
    \coordinate (H5) at (3.5,-0.866);
    \coordinate (H6) at (4.5,0.866);
    \coordinate (H7) at (5,1.732);
    \coordinate (H8) at (5,0);
    \coordinate (H9) at (5,-1.732);
    \coordinate (H10) at (4.5,-0.866);

    \foreach \start/\andd in {A/B1, A/B2, A/B3, B1/C1, B1/C2, B3/C3, B3/C4, B2/C5, B2/C6, C1/D1, C6/D1, C2/D2, C3/D2, C4/D3, C5/D3, C2/E1, C1/E2, C6/E3, C5/E4, C4/E5, C3/E6, D1/H1, E2/H2, E3/H3, H1/H4, H2/H4, H1/H5, H3/H5, H4/H6, H6/H7, H5/H10, H10/H8, H10/H9, H8/H6} {
        \draw (\start) -- (\andd);
    }

    \node[atomA] (A1) at (A) {};
    \node[atomB] at (B1) {};
    \node[atomB] at (B2) {};
    \node[atomB] at (B3) {};
    \node[atomA] at (C1) {};
    \node[atomA] at (C2) {};
    \node[atomA] at (C3) {};
    \node[atomA] at (C4) {};
    \node[atomA] at (C5) {};
    \node[atomA] at (C6) {};
    \node[atomB] at (D1) {};
    \node[atomB] at (D2) {};
    \node[atomB] at (D3) {};
    \node[atomB] at (E1) {};
    \node[atomB] at (E2) {};
    \node[atomB] at (E3) {};
    \node[atomB] at (E4) {};
    \node[atomB] at (E5) {};
    \node[atomB] at (E6) {};

    \node[atomA] at (H1) {};
    \node[atomA] at (H2) {};
    \node[atomA] at (H3) {};
    \node[atomB] at (H4) {};
    \node[atomB] at (H5) {};
    \node[atomA] at (H6) {};
    \node[atomB] at (H7) {};
    \node[atomB] at (H8) {};
    \node[atomB] at (H9) {};
    \node[atomA] at (H10) {};
    
    \def\offset{0.11}
    
    \draw[vector] ([shift={( 0.5*\offset, 0.866*\offset)}] A) -- ([shift={(-0.5*\offset, -0.866*\offset)}] B1) node[pos=0.6, left] {$t$};
    
    \draw[vector] ([shift={( 0.75*\offset, 0.433*\offset)}] A) -- ([shift={(-0.75*\offset, -0.433*\offset)}] C1) node[pos=0.5, below] {$\quad t_2e^{i\sigma\varphi}$};
    
    \draw[vector] ([shift={( 0.75*\offset, -0.433*\offset)}] A) -- ([shift={(-0.75*\offset, 0.433*\offset)}] C6) node[pos=0.4, right] {$\, t_2e^{-i\sigma\varphi}$};
    
    \draw[vector] ([shift={( 0*\offset, -\offset)}] A) -- ([shift={(0*\offset, \offset)}] C5) node[pos=0.8, right] {$t_2e^{i\sigma\varphi}$};
    
    \draw[vector] ([shift={( -0.5*\offset, -0.866*\offset)}] A) -- ([shift={(0.5*\offset, 0.866*\offset)}] D3) node[pos=0.5, left] {$t_3$};
    
    \draw[vector] ([shift={( -1*\offset, -0.346*\offset)}] A) -- ([shift={(1*\offset, 0.346*\offset)}] E5) node[pos=0.6, above] {$t_4$};

    \draw[vector] ([shift={( 0.75*\offset, 0.433*\offset)}] H4) -- ([shift={(-0.75*\offset, -0.433*\offset)}] H7) node[pos=0.8, left] {$t_2e^{-i\sigma\varphi}$};

    \draw[vector] ([shift={( 0.75*\offset, -0.433*\offset)}] H4) -- ([shift={(-0.75*\offset, 0.433*\offset)}] H8) node[pos=0.5, right] {$\, t_2e^{i\sigma\varphi}$};

    \draw[vector] ([shift={( 0*\offset, -\offset)}] H4) -- ([shift={(0*\offset, \offset)}] H5) node[pos=0.8, right] {$t_2e^{-i\sigma\varphi}$};

    \node[above left] at (A) {$A$};
    \node[above left] at (H4) {$B\,\,$};
    \node[labelStyle, right] at (C2) {$+M$};
    \node[labelStyle, right] at (E1) {$-M$};
    
    \end{tikzpicture}}\label{fig:extended-kane-mele-model-drawing}}

    \subfigure[]{
    \includegraphics[width=0.8\linewidth]{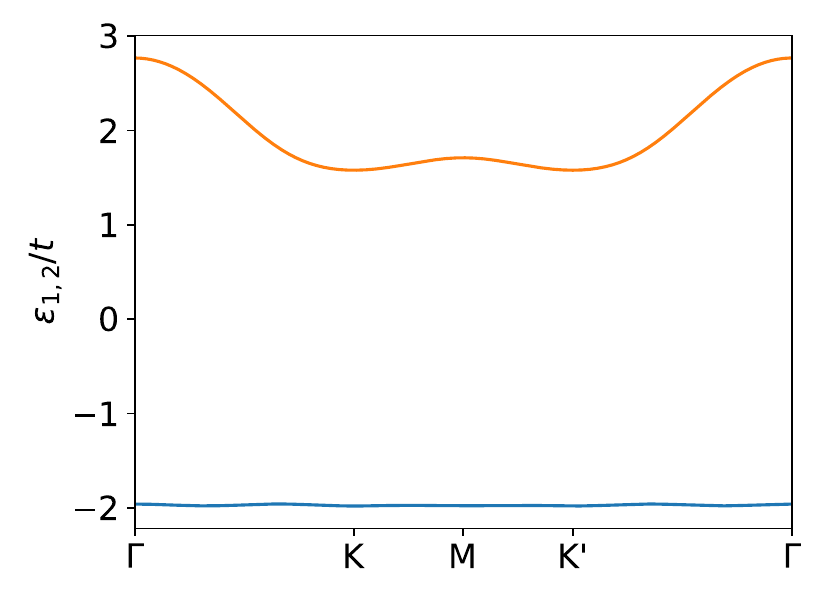}
    \label{fig:kane-mele-flat-band}
    }
    
    \caption{\textbf{(a)} Sketch of the extended Kane-Mele model on a honeycomb lattice with sublattice $A$ (filled circles) and sublattice $B$ (empty circles) with red arrows indicating the hopping amplitudes, similar to the illustration in Ref.\ \cite{lau2022universal}. \textbf{(b)} Energy bands for $t_2 = 0.349t$, $t_3 = -0.264t$, $t_4=0.026t$, $\varphi = 1.377$, and $M = 0$. In this case, the Kane-Mele model hosts a nearly dispersionless band.}
    
\end{figure}
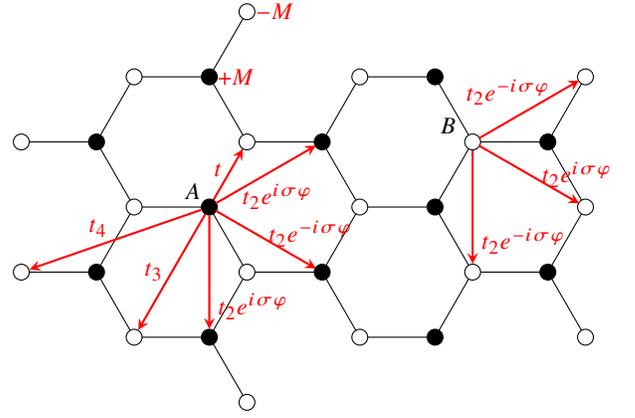
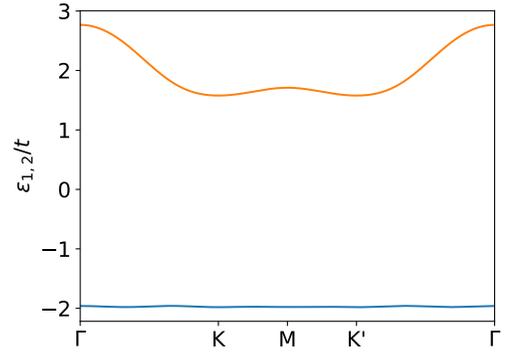 \\
\indent Let us perform a spatial Fourier transformation of the extended Kane-Mele Hamiltonian. The lattice vectors we use for the description of the honeycomb lattice in Fig.~\ref{fig:extended-kane-mele-model-drawing} are
\begin{align}
    \vv{a}_1 = \frac{a}{2} \svec{3 \\ -\sqrt{3}}\,, \qquad \vv{a}_2 = a\svec{0\\ \sqrt{3}} \,,
\end{align}
where $a$ represents the lattice constant. The corresponding reciprocal lattice vectors are given by
\begin{align}
    \vv{b}_1 = \frac{2\pi}{a} \svec{2/3\\ 0}\,, \qquad \vv{b}_2 = \frac{2\pi}{a} \svec{1/3\\ 1/\sqrt{3}} \,.
    \label{eq:reciprocal-lattice-vectors}
\end{align}
Furthermore, we set the basis vectors which describe the positions of the two atoms $A$ and $B$ in each unit cell to be
\begin{align}
    \vv{r}_A = \svec{0\\0}\,, \qquad \vv{r}_B = \frac{a}{2}\svec{1\\ \sqrt{3}} \,.
\end{align}
Summations over $n$th nearest neighbors in the Hamiltonian~\eqref{eq:extended-kane-mele-model} can be expressed by
\begin{align}
    \sum_{\kla{i,j}_n} f(\vv{R}_i, \vv{R}_j) = \sum_{i} \sum_{j=1}^{m_n} f(\vv{R}_i, \vv{R}_i + \vvv{\delta}^{(n)}_j) \,,
\end{align}
where $m_n$ represents the number of $n$th nearest neighbors. This allows us to perform a Fourier transformation such that we can obtain an expression for the Bloch Hamiltonian $H(\vv{k})$ which is needed to obtain quantum geometrical and topological information of the system. We define
$
    \mathcal{C}^\dagger(\vv{k}) \coloneqq ((c^\dagger_{A\uparrow}(\vv{k}), c^\dagger_{B\uparrow}(\vv{k}), c^\dagger_{A\downarrow}(\vv{k}), c^\dagger_{B\downarrow}(\vv{k}))
$ 
and after some calculation, it turns out that $H(\vv{k})$ has block-diagonal form and $H_0$ can be expressed as
\begin{align}
    H_0 = \sum_{\vv{k}} \mathcal{C}^\dagger(\vv{k}) \underbrace{\matr{H_\uparrow(\vv{k}) & 0\\ 0 & H_\downarrow(\vv{k})}}_{=H(\vv{k})} \mathcal{C}(\vv{k})\,,
\end{align}
whereby $H_\uparrow(\vv{k}) = H^T_\downarrow(-\vv{k})$ represents the time-reversal symmetric Bloch Hamiltonian
\begin{align}
    H_\sigma(\vv{k}) = \matr{M + g_+(\sigma, \vv{k}) & f_1(\vv{k}) + f_3(\vv{k}) + f_4(\vv{k}) \\ f_1^\ast(\vv{k}) + f_3^\ast(\vv{k}) + f_4^\ast(\vv{k}) & -M + g_-(\sigma,\vv{k})} \,.
    \label{eq:single-particle-hamiltonian-kane-mele}
\end{align}
Here, the functions $f,g$ represent the different contributions that come out of the different nearest-neighbor sums. With $t_1 \equiv t$ it turns out that the off-diagonal elements are determined by 
\begin{align}
    f_n(\vv{k}) &= t_n \sum_j e^{-i\vv{k} \cdot \vvv{\delta}^{(n)}_j}\,,
\end{align}
and the diagonal elements are determined by
\begin{align}
    g_\pm(\sigma,\vv{k}) &= t_2 \sum_{j} e^{-i\klr{\vv{k} \cdot \vvv{\delta}^{(2)}_j \pm (-1)^j \sigma \varphi}} \,.
\end{align}
Using
\mbox{$
    \Psi_{\vv{k}} = (c_{\alpha \uparrow}(\vv{k}), c_{\alpha \downarrow}(\vv{k}), c^\dagger_{\alpha \uparrow}(-\vv{k}), c^\dagger_{\alpha \downarrow}(-\vv{k}))
$}
as our Nambu spinor where $\alpha = A,B$ denotes the sublattice, the BdG Hamiltonian defined in Eq.~\eqref{eq:bdg-hamiltonian} takes the following form~\cite{lau2022universal}:
\begin{widetext}
\begin{align}
    \mathcal{H}(\vv{k}) = \matr{H_\uparrow(\vv{k}) - \mu\mathbbm{1} & 0 & \Delta_{\uparrow\uparrow} & \Delta_{\uparrow\downarrow} \\ 0 & H_\downarrow(\vv{k}) - \mu\mathbbm{1} & \Delta_{\downarrow\uparrow} & \Delta_{\downarrow\downarrow} \\ \Delta^\dagger_{\uparrow\uparrow} & \Delta^\dagger_{\downarrow\uparrow} & -H^T_\uparrow(-\vv{k}) + \mu\mathbbm{1} & 0\\ \Delta^\dagger_{\uparrow\downarrow} & \Delta^\dagger_{\downarrow\downarrow} & 0 & -H^T_\downarrow(-\vv{k}) + \mu\mathbbm{1}} \,.
\end{align}
Here, $\mu$ represents the chemical potential and $\Delta_{\sigma\sigma'}$ represent $(2 \times 2)$ the gap function matrices. For simplicity, we confine ourselves to spin-singlet pairing \footnote{In principle, we can perform the calculations for any unconventional superconducting state on a hexagonal lattice, including spin-triplet pairing.}, implying the conditions $\Delta_{\downarrow\downarrow} = \Delta_{\uparrow\uparrow} = 0$ and $\Delta_{\uparrow\downarrow} = -\Delta_{\downarrow\uparrow}^T \equiv \Delta$~\cite{lau2022universal}. Consequently, the Bogoliubov-de Gennes Hamiltonian can always be unitarily transformed into block‑diagonal form 
\begin{align}
    \tilde{\mathcal{H}}(\vv{k}) &= \matr{H_\uparrow(\vv{k}) - \mu\mathbbm{1} & \Delta & 0 & 0 \\ \Delta^\dagger & -H^T_\downarrow(-\vv{k}) + \mu\mathbbm{1} & 0 & 0 \\ 0 & 0 & H_\downarrow(\vv{k}) - \mu\mathbbm{1} & -\Delta^T \\ 0 & 0 & -\Delta^\ast & -H_\uparrow^T(-\vv{k}) + \mu\mathbbm{1} } \eqqcolon \matr{\mathcal{H}_\uparrow(\vv{k}) & 0\\ 0 & \mathcal{H}_\downarrow(\vv{k})} \,.
\end{align}
\end{widetext}
Thus, the BdG Hamiltonian reduces to smaller BdG Hamiltonians $\mathcal{H}_\sigma(\vv{k})$ associated to the spin~$\sigma$ and it is sufficient to look at one of the spin-dependent BdG Hamiltonians only, e.g.,
\begin{align}
    \mathcal{H}_\uparrow(\vv{k}) = \matr{H_\uparrow(\vv{k}) - \mu\mathbbm{1} & \Delta \\ \Delta^\dagger & -H_\downarrow^T(-\vv{k}) + \mu\mathbbm{1}} \,,
    \label{eq:spin-up-bdg-hamiltonian}
\end{align}
and calculate the superfluid weight at zero temperature associated to this Hamiltonian with the procedure described in Sec.~\ref{sec:calculation-superfluid-weight-functional}. 

In particular, to obtain the superfluid weight, we calculate the conventional contribution using the formula in Eq.~\eqref{eq:superfluid-weight-conv-def}
which depends on the Bogoliubov coefficients and the curvature of the energy bands only. Moreover, to keep the numerical implementation as simple as possible, we employ Eq.~\eqref{eq:superfluid-weight-geom} to determine the geometrical contribution which depends on the second derivatives of the diagonalized single-particle Hamiltonian, i.e., its energy bands $\varepsilon_{\vv{k}}$, and quasiparticle eigenvalues $E_{\vv{k}n}$. To estimate the functional contribution, we refer to Eq.~\eqref{eq:functional-superfluid-weight-wrt-skew-matrices} 
which depends on the single-particle Hamiltonian, the single-particle eigenstates, the Bogoliubov coefficients, and the pairing potential which is encoded within the inverse of the matrix $\hat{M}$. For all pairing mechanisms we have considered for the extended Kane-Mele model, our pairing potential factorizes such that we can apply the Sherman-Morrison-Woodbury formula (cf.\ Appendix~\ref{sec:woodbury}). In particular, for the conventional $s$-wave and chiral $d$-wave superconducting states we refer to Eqs.~\eqref{eq:conventional-M} and \eqref{eq:unconventional-M}, respectively.

The Kane-Mele model \eqref{eq:extended-kane-mele-model} is a tight-binding Hamiltonian defined on a honeycomb lattice with point group $C_{6v}$ \cite{platt2013functional}. In Table~\ref{tab:C6v_character_table}, we have summarized the character table with the corresponding basis functions.  
Since the pair spin-wave-function of a singlet is antisymmetric with respect to an interchange of the spin indices, the gap function can be expressed as \cite{sigrist1991phenomenological,mineev1999introduction}
\begin{align}
    \Delta_\alpha(\vv{k}) = g_\alpha(\vv{k}) i\sigma_y\,, \qquad g_\alpha(\vv{k}) = g_\alpha(-\vv{k}) \,,
\end{align}
where $g_\alpha$ is determined by the symmetry of the system. If~$\Gamma$ denotes an irreducible representation of $C_{6v}$ with dimension $d_\Gamma$ and $\psi_i^{\Gamma}$ are the simplest basis functions even in $\vv{k}$, respecting the symmetry of the system, the coefficient $g_\alpha$ is~\cite{mineev1999introduction}
\begin{align}
    g_\alpha(\vv{k}) = \sum_{i=1}^{d_\Gamma} \Delta_\alpha^i \psi_i^{\Gamma}(\vv{k}) \,,
    \label{eq:form-factor-definition}
\end{align}
and the pairing potential takes the form
\begin{align}
    U(\vv{k},\vv{k}') = U_0 \sum_{i=1}^{d_\Gamma} \psi_i^{\Gamma}(\vv{k}') \bar{\psi}_i^\Gamma(\vv{k}) \,.
    \label{eq:definition-pairing-potential}
\end{align}
Here, $\klg{\Delta_\alpha^i}$ represent the $d_\Gamma$ order parameters of the $\alpha$th band. 

\begin{table}[t!]
\caption{Character table for point group $C_{6v}$ based on \cite{powell2010symmetry,black2014chiral}. The irreps with even basis functions ($A_1,A_2,E_2$) correspond to singlet superconducting states, which are the focus of the following discussion, while those with odd basis functions ($B_1,B_2,E_1$) describe triplet states and are therefore not considered further. Note that the irreps $E_1$ and $E_2$ are two-dimensional whereas the others are one dimensional.}\label{tab:C6v_character_table}
\renewcommand{\arraystretch}{1.2}
\begin{ruledtabular}
\begin{tabular}{cccccccl}
\textrm{Irreps} & \textrm{$E$} & \textrm{$2C_6$} & $2C_3$ & $C_2$ & $3\sigma_v$ & $3\sigma_d$ & Basis function(s) \\ 
\colrule
$A_1$ & $\phantom{-}1$ & $\phantom{-}1$ & $\phantom{-}1$ & $\phantom{-}1$ & $\phantom{-}1$ & $\phantom{-}1$ & $1,\quad k_x^2+k_y^2$ \\
        $A_2$ & $\phantom{-}1$ & $\phantom{-}1$ & $\phantom{-}1$ & $\phantom{-}1$ & $-1$ & $-1$ & $k_xk_y(k_x^2 - 3k_y^2)(k_y^2 - 3k_x^2)$ \\
        $B_1$ & $\phantom{-}1$ & $-1$ & $\phantom{-}1$ & $-1$ & $\phantom{-}1$ & $-1$ & $k_x (k_x^2 - 3k_y^2)$ \\
        $B_2$ & $\phantom{-}1$ & $-1$ & $\phantom{-}1$ & $-1$ & $-1$ & $\phantom{-}1$ & $k_y (k_y^2 - 3k_x^2)$ \\
        $E_1$ & $\phantom{-}2$ & $\phantom{-}1$ & $-1$ & $-2$ & $\phantom{-}0$ & $\phantom{-}0$ & $(k_x,k_y)$ \\
        $E_2$ & $\phantom{-}2$ & $-1$ & $-1$ & $\phantom{-}2$ & $\phantom{-}0$ & $\phantom{-}0$ & $(k_x^2 - k_y^2,k_xk_y)$ \\
\end{tabular}
\end{ruledtabular}
\end{table}
%

\subsection{Self-consistent equation}

For the following evaluations of the superfluid weights, we use the general formulas Eqs.~\eqref{eq:superfluid-weight-conv-def},~\eqref{eq:superfluid-weight-geom}, and~\eqref{eq:functional-superfluid-weight-wrt-skew-matrices}. This implies that the superconducting gap function is not constrained to being proportional to the identity matrix.
Recall that the BdG Hamiltonian has been diagonalized within two steps. 
First, we diagonalized the single-particle Bloch Hamiltonian $H(\vv{k})$ using the matrix $S(\vv{k})$ (which contains all the information about the quantum geometry) [cf.\ Eq.~\eqref{eq:diagonalization-single-particle}], and in a second step we diagonalized the BdG Hamiltonian using the matrix $W_{\vv{k}}(\vv{q})$ (which contains the Bogoliubov coefficients for $\vv{q} = 0$) [cf.\ Eq.~\eqref{eq:diagonalization-bdg-hamiltonian}]. Therefore, the $2N_\mathrm{B}$ eigenvectors of the BdG Hamiltonian \eqref{eq:mean-field-hamiltonian-for-nambu} in Nambu spinor basis can be expressed as the columns of
\begin{align}
    \svec{v_{n,+}(\vv{q}; \vv{k})\\ v_{n,-}(\vv{q};\vv{k})} = \kle{\matr{S(\vv{k}-\vv{q}) & 0\\ 0 & S^\ast(-\vv{k}-\vv{q})} \cdot W_{\vv{k}}(\vv{q})}_{\cdot, n} \,,
    \label{eq:eigenvectors-bdg-hamiltonian}
\end{align}
where $n = 1,\hdots, 2N_\mathrm{B}$. We introduce new creation operators $\klg{\tilde{\psi}_{n\vv{k}}; n = 1,\hdots,2N_\mathrm{B}}$, defined by \cite{lau2022universal}
\begin{align}
    \psi^\dagger_{\vv{k}\alpha} &= \sum_{n=1}^{2N_\mathrm{B}} v^\ast_{n\alpha,+}(\vv{q};\vv{k}) \tilde{\psi}^\dagger_{n\vv{k}} \,, \quad
    \psi_{-\vv{k}\alpha} = \sum_{n=1}^{2N_\mathrm{B}} v^\ast_{n\alpha,-}(\vv{q};\vv{k}) \tilde{\psi}^\dagger_{n\vv{k}} \,.
\end{align}
It can be easily checked that these also fulfill fermionic commutation relations and, using these creation operators, the BdG Hamiltonian \eqref{eq:bdg-hamiltonian} becomes 
\begin{align}
    \mathcal{H}_{\mathrm{BdG}}(\vv{k}, \vv{q}) = \sum_{\vv{k},n} E_{\vv{k}n}(\vv{q}) \tilde{\psi}^\dagger_{n\vv{k}} \tilde{\psi}_{n\vv{k}}\,,
\end{align}
and we can rewrite the self-consistency equation \eqref{eq:self-consistent-equation} as \cite{lau2022universal}
\begin{align}
    \Delta_{\alpha\beta}(\vv{q};\vv{k}') = & \frac{1}{V} \sum_{\vv{k},n}  U(\vv{k},\vv{k}') v_{n\alpha,+}(\vv{q};\vv{k}) v^\ast_{n\beta,-}(\vv{q};\vv{k})\notag\\
    &\quad\quad\times\klr{1 - n_{\mathrm{F}}(E_{\vv{k}n}(\vv{q}))}\,,
\end{align}
where $n_{\mathrm{F}}$ denotes the Fermi-Dirac function. If the gap function $\Delta(\vv{q};\vv{k})$ can be represented by a diagonal matrix (intraband pairing), then it is sufficient to solve the self-consistent equation for the $\alpha = \beta$ cases as it should also satisfy automatically the self-consistency equations for the $\alpha \neq \beta$ cases. Since we consider spin singlets only, we can insert Eq.~\eqref{eq:form-factor-definition} into the above equation and obtain
\begin{align}
    \sum_{i=1}^{d_\Gamma} \Delta_{\alpha}^i(\vv{q}) \psi_i^{\Gamma}(\vv{k}') = & \frac{U_0}{V} \sum_{i=1}^{d_\Gamma} \psi_i^{\Gamma}(\vv{k}') \sum_{\vv{k},n} \bar{\psi}_i^{\Gamma}(\vv{k}) v_{n\alpha,+}(\vv{q};\vv{k})\notag\\
    &\times v^\ast_{n\alpha,-}(\vv{q};\vv{k}) \klr{1 - n_{\mathrm{F}}(E_{\vv{k}n}(\vv{q}))} \,.
\end{align}
We utilize the linear independence of the basis functions and end up with a $(d_\Gamma N_{\mathrm{B}})$-dimensional fix-point equation we need to solve:
\begin{align}
    \Delta^i_{\alpha}(\vv{q}) = & \frac{U_0}{V} \sum_{\vv{k},n} \bar{\psi}_i^{\Gamma}(\vv{k}) v_{n\alpha,+}(\vv{q};\vv{k}) v^\ast_{n\alpha,-}(\vv{q};\vv{k}) \notag\\
    &\quad\quad\times\klr{1 - n_{\mathrm{F}}(E_{\vv{k}n}(\vv{q}))} \,.
\end{align}
Note that if the gap function is not of intraband type, then a new ansatz for the matrix gap function is necessary and we may end up with a $(d_\Gamma N_{\mathrm{B}}^2)$-dimensional fix-point equation. On the other hand, if the intraband pairing condition is fulfilled, then the order parameters are independent of the band indices and we need to solve a $d_\Gamma$-dimensional fix-point equation.
\subsection{Conventional $s$-wave superconductivity}\label{sec:kane-mele-s-wave}
\begin{figure*}[t]
    \centering
    \subfigure[]{
        \includegraphics[width=0.32\textwidth]{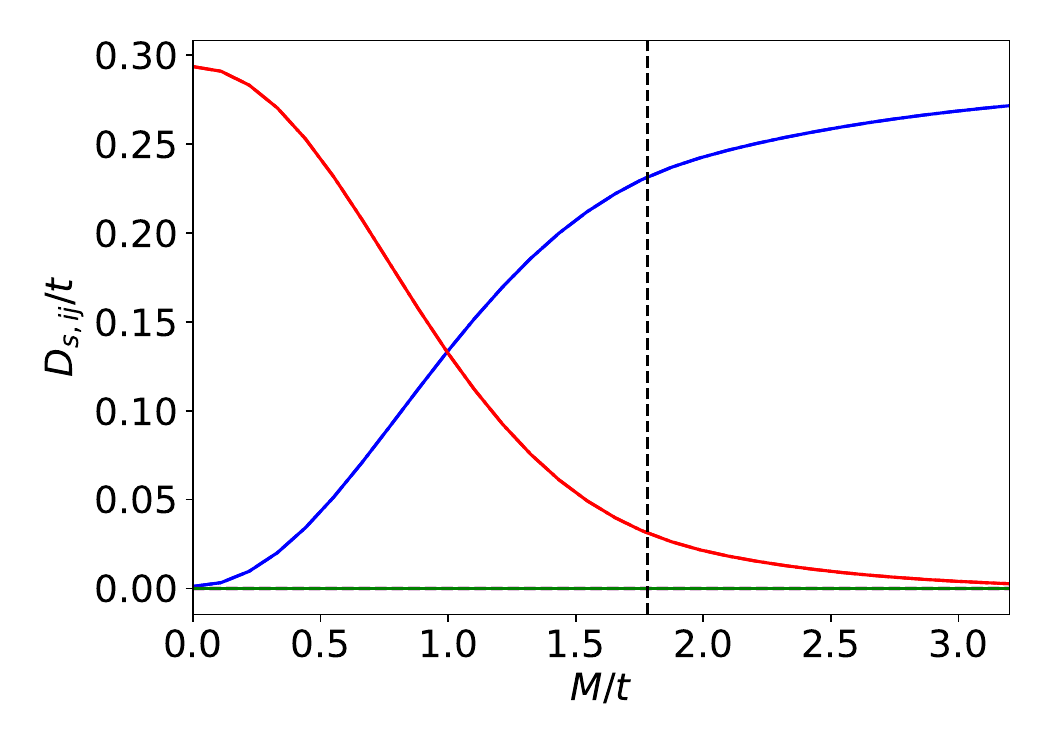}
        \label{fig:superfluid-weight-along-M-axis-s-wave}
    }
    \subfigure[]{
        \includegraphics[width=0.32\textwidth]{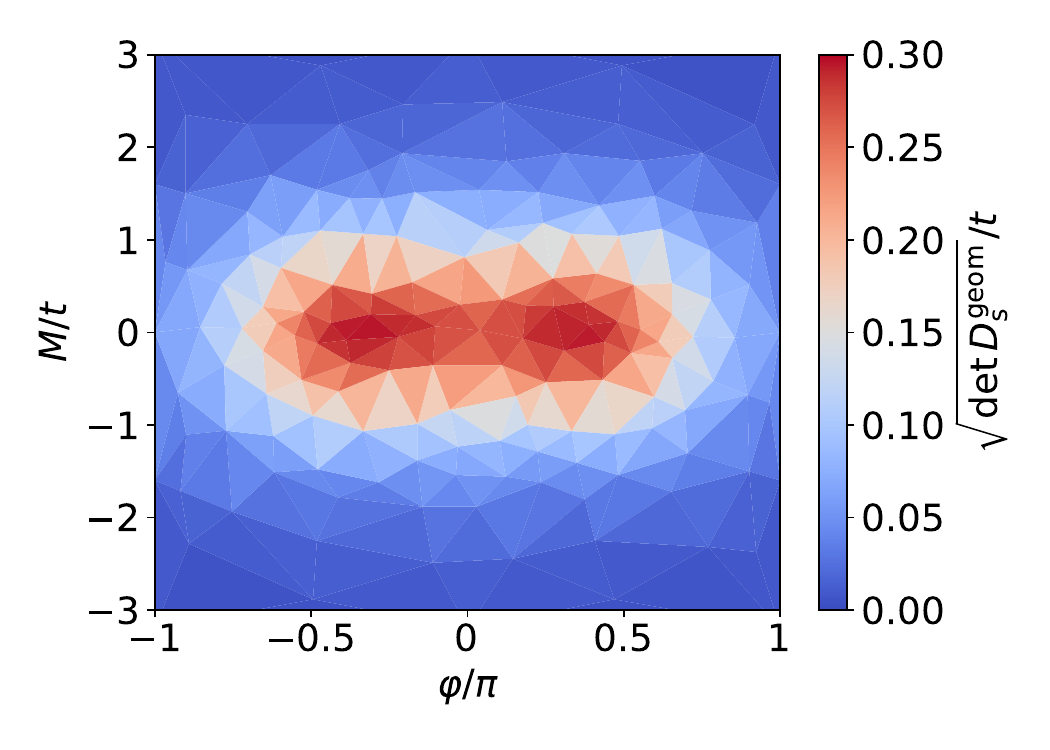}
       \label{fig:superfluid-weight-along-M-phi-plane-s-wave}
   }
    \caption{Kane-Mele Hamiltonian for fixed $t_2 = 0.349t$, $t_3 = -0.264t$, and $t_4 = 0.026t$ with $s$-wave superconducting order parameter whereas the Brillouin zone has been discretized by a $20 \times 20$ grid. In both plots, the numerical errors are of order $\le 1\%$. \textbf{(a)}~For fixed interaction strength $U_0 = 3t$, the blue, red, and green curves represent the conventional, geometrical, and functional superfluid weight \mb{along the $M$-axis for fixed $\varphi = 1.377$}. While the solid colored lines represent the diagonal elements, the dashed colored lines (which are zero) represent the off-diagonal elements. The black dashed vertical line indicates the change of Chern number. 
    \textbf{(b)}~Geometrical superfluid weight for fixed interaction strength $U_0 = 3t$ plotted against the hopping phase and the staggered onsite potential.}
\end{figure*}
For comparison, we start with the discussion of the conventional $s$-wave superconducting state which possesses the full point symmetry of the lattice and belongs to the irrep $A_1$ given in Table~\ref{tab:C6v_character_table}. The advantage here is that we only need to solve a two-dimensional fix-point equation for the determination of the order parameter. Moreover, we can compare and gauge our results with Ref.~\cite{lau2022universal}. \\
\indent As we assume intraband pairing only, according to the previous section we need to solve the self-consistent equation
\begin{align}
    \Delta_\alpha(\vv{q}) = \frac{U_0}{V} \sum_{\vv{k},n} v_{n\alpha,+}(\vv{q};\vv{k}) v^\ast_{n\alpha,-}(\vv{q};\vv{k}) \klr{1 - n_{\mathrm{F}}(E_{\vv{k}n}(\vv{q}))}
    \label{eq:self-consistency-equation-useful-s-wave}
\end{align}
for each band $\alpha = 1,2$ where $U_0$ represents the interaction strength. In the specific case $M = 0$ we can drop the orbital dependency of the order parameter due to symmetry reasons and the fix-point equation reduces to a one-dimensional one. Here, $n_{\mathrm{F}}(E_{\vv{k}n}(\vv{q}))$ represents the Fermi function, which for $T\to 0$ reduces to a step function.
For the precise and fast determination of the order parameter, we use the library \texttt{scipy.optimize} and its \texttt{fsolve} method~\cite{scipy-optimize-fsolve}. 
As an initial condition for the optimization process, we have set the interaction strength $U_0$ as we expect the order parameter to be of same order of magnitude. Some sample plots for $M = 0$ are given in Fig.~\ref{fig:s-wave-self-consistent-equations} of Appendix~\ref{sec:appendix-fig-s-wave}.
These fix points are consistent with the results presented in Ref.\ \cite{lau2022universal}. \\
\indent Hence, we can continue the determination of the superfluid weight along the $M$-axis for the fixed hopping phase $\varphi = 1.377$. This setting is ideal because, at 
$M=0$, the Kane-Mele model hosts an almost flat band [cf.\ Fig.~\ref{fig:kane-mele-flat-band}], allowing us to probe the influence of quantum geometry. 
The result is shown in Fig.~\ref{fig:superfluid-weight-along-M-axis-s-wave}. We make three observations. First, as stated in Ref.~\cite{lau2022universal}, the superfluid weight is proportional to the unity matrix, i.e., the off-diagonal elements are zero and the diagonal elements have the same value. Moreover, as expected, we see that in the topologically nontrivial flat-band limit, $M=0$, the geometrical contribution dominates whereas in the topological trivial dispersing bands limit, $M = 3.2t$, the conventional contribution dominates. We also observe that the functional superfluid weight is zero which agrees with our expectations because the gap function does not break TRS and the imaginary part of the order parameter is zero. In addition, in Fig.~\ref{fig:superfluid-weight-along-M-phi-plane-s-wave} we see that when compared to the topological phase diagram of the Haldane model \cite{haldane1988model}, the presence of nontrivial single-particle topology seems to result in an increased value for the geometrical superfluid weight. This is consistent with the Wirtinger inequality which states that the Chern number represents a lower bound of the quantum metric~\cite{wirtinger1936determinantenidentitat,ozawa2021relations,mera2021kahler,mera2022relating}.

\subsection{Chiral $d$-wave superconductivity}\label{sec:kane-mele-d-wave}

Let us now discuss chiral $d$-wave superconductivity. Many works on unconventional superconductivity in a honeycomb lattice consider the possibility of chiral $d$-wave pairing, which corresponds to the superconducting state that belongs to the two-dimensional irrep $E_2$ in Table~\ref{tab:C6v_character_table}. In general, it is a spin singlet characterized by a linear combination of two order parameters per orbital \cite{black2014chiral,black2014chiral_mott}. 
The phase of the chiral $d$-wave state winds around the Brillouin zone center twice which defines a nonzero topological invariant related to the number of edge modes. Because these modes are propagating only in one direction but not in the opposite direction they are called chiral (cf. Ref.~\cite{black2014chiral}).

The form factors $\psi_i^\Gamma(\vv{k})$ are determined by using the basis functions such that they obey the translational symmetry of the reciprocal lattice and have as few nodes as possible. By using the reciprocal basis vectors of the honeycomb lattice we have stated in Eq.~\eqref{eq:reciprocal-lattice-vectors}, it turns out that they are given by~\cite{platt2013functional,black2014chiral}
\begin{align}
    \psi^{E_2}_{x^2-y^2}(\vv{k}) =& \frac{1}{\sqrt{6}} \Big( 2\cos\kle{\sqrt{3}ak_y} - \cos\kle{\klr{\sqrt{3}k_y - 3k_x}a/2} \nonumber \\
    &- \cos\kle{\klr{\sqrt{3}k_y + 3k_x}a/2} \Big)\,, \label{eq:basis-functions-chiral-d-wave-hex-1}\\[8pt]
    \psi^{E_2}_{xy}(\vv{k}) =& \frac{1}{\sqrt{2} }\Big(\cos\kle{\klr{\sqrt{3}k_y - 3k_x}a/2} \nonumber \\
    &- \cos\kle{\klr{\sqrt{3}k_y + 3k_x}a/2} \Big)\,.
    \label{eq:basis-functions-chiral-d-wave-hex-2}
\end{align}
Because the system favors as few nodes as possible, it can be shown with the help of the Ginzburg-Landau phenomenology (cf.\ e.g.\ \cite{black2014chiral_mott}) that a complex combination of the form factors is usually favored for the $d$-wave superconducting state. In particular, we expect the order parameters associated with the $x^2-y^2$ and $xy$ form factors to have the same magnitude and to exhibit a phase shift of $\pi/2$ \cite{black2014chiral}. 

\begin{figure}[b!]
    \centering
    \includegraphics[width=\columnwidth]{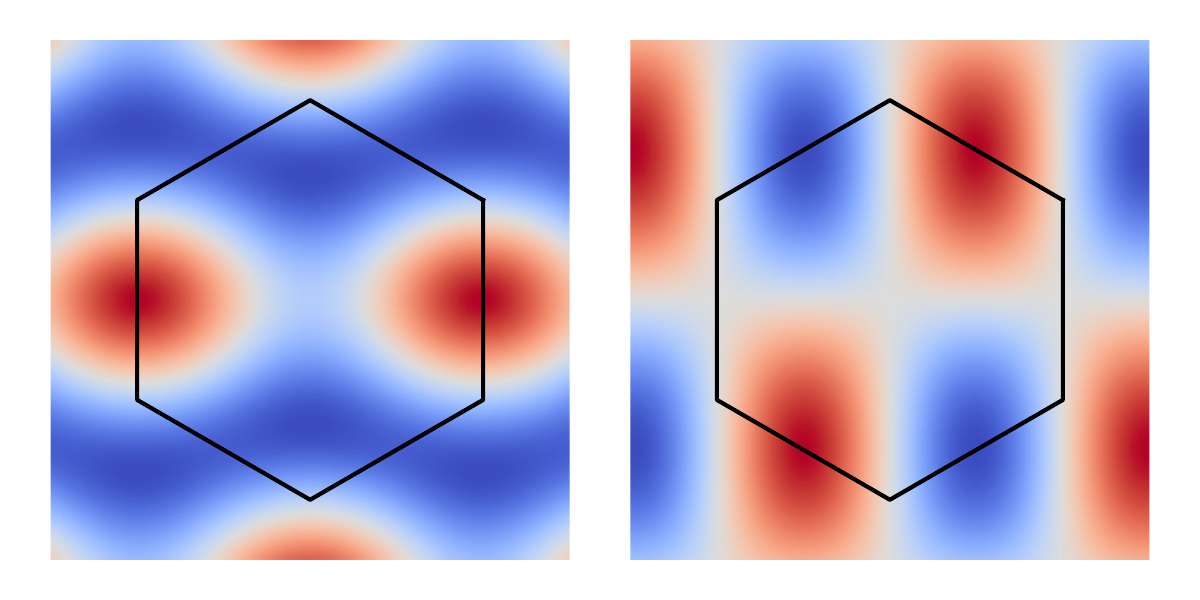}
    \caption{Real and imaginary components of the chiral $d$-wave superconducting gap function, corresponding to the form factors given in Eqs.~\eqref{eq:basis-functions-chiral-d-wave-hex-1} (left) and \eqref{eq:basis-functions-chiral-d-wave-hex-2} (right), respectively \cite{black2014chiral}. The black lines indicate the hexagonal Brillouin zone.}
    \label{fig:dwave-basis-functions}
\end{figure}
\begin{figure*}[t!]
    \centering
    \subfigure[]{
        \includegraphics[width=0.32\textwidth]{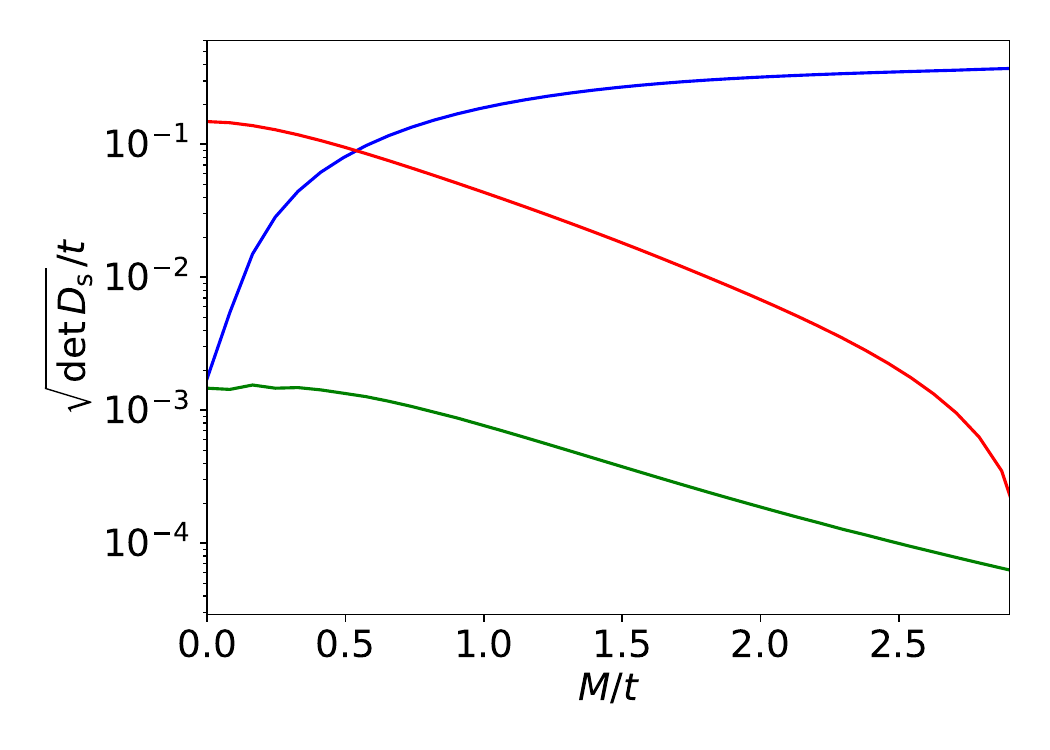}
        \label{fig:superfluid-weight-along-M-axis-d-wave}
    }
    \subfigure[]{
        \includegraphics[width=0.32\textwidth]{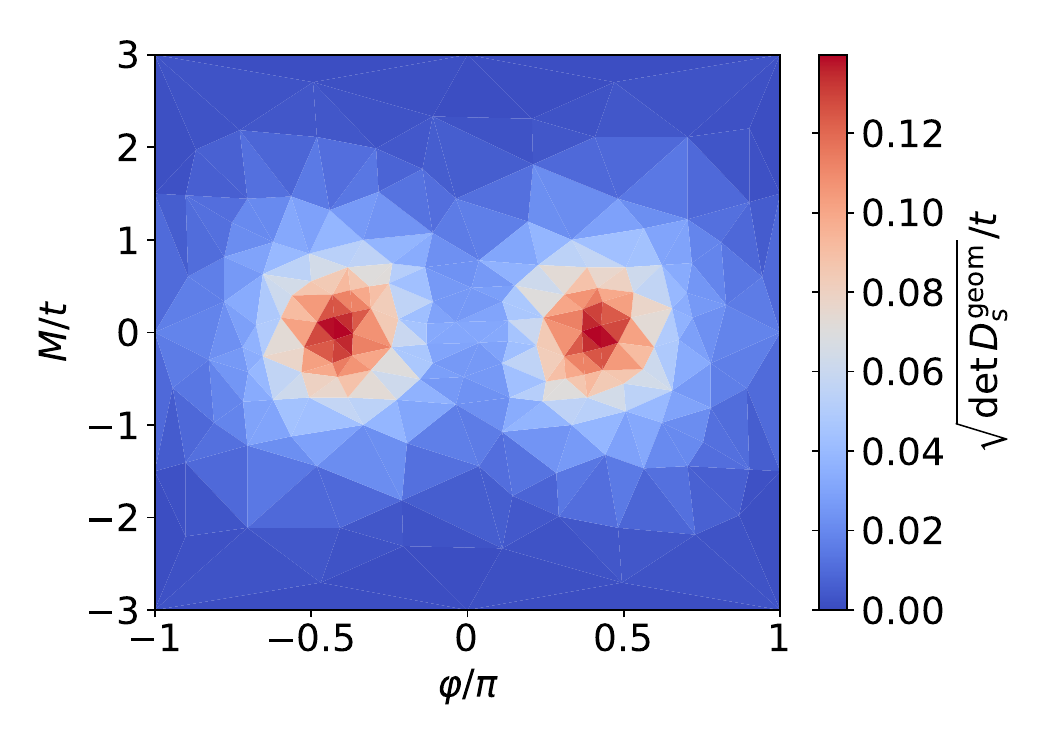}
        \label{fig:geom-weight-dwave-plane}
    }
    \subfigure[]{
        \includegraphics[width=0.32\textwidth]{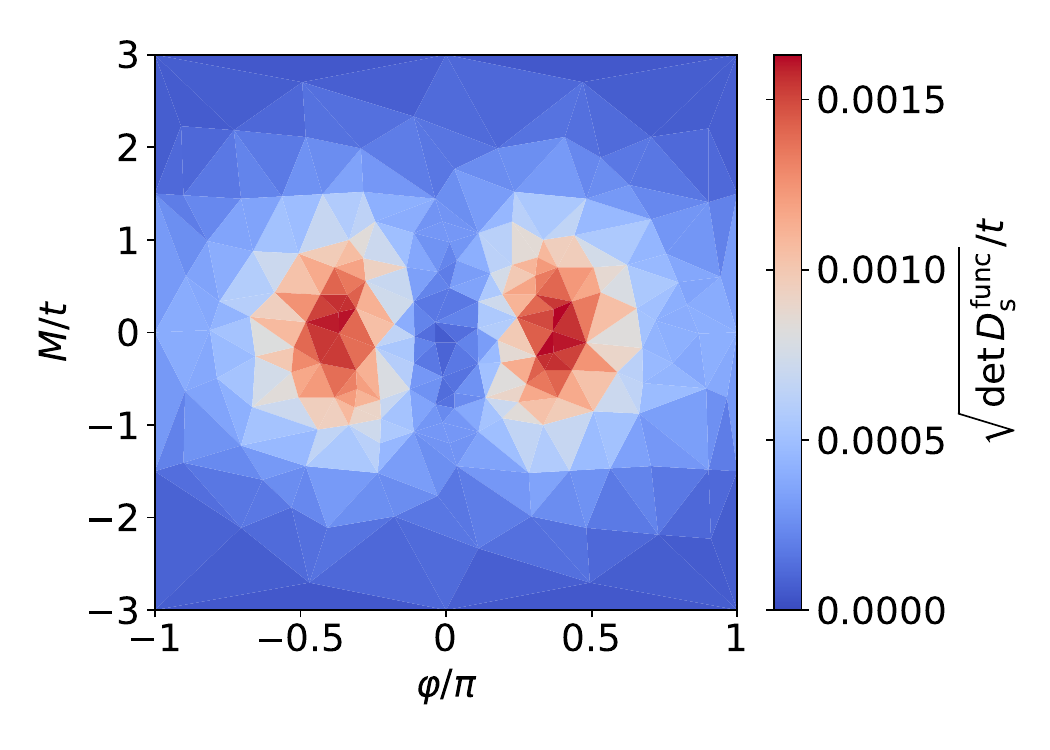}
        \label{fig:func-weight-dwave-plane}
    }
    \caption{Kane-Mele Hamiltonian for fixed $t_2 = 0.349t$, $t_3 = -0.264t$, and $t_4 = 0.026t$ with chiral $d$-wave superconducting order parameter whereas the Brillouin zone has been discretized by a $20 \times 20$ grid. In all of these plots, the numerical errors are of order $\le 1\%$.  \textbf{(a)} Logarithmic plot of the conventional (blue), geometrical (red) and functional (green) contributions to the superfluid weight \mb{along the $M$-axis for fixed $\varphi = 1.377$} with an interaction strength of $U_0 = 3t$. Note that the functional superfluid weight has a negative contribution to the total superfluid weight.
    \textbf{(b)} Geometrical and \textbf{(c)} functional superfluid weight for fixed interaction strength $U_0 = 3t$ plotted against the hopping phase and the staggered onsite potential. }
    \label{fig:superfluid-weight-overview-dwave}
\end{figure*} 

Therefore, we make the following ansatz as the gap function~$\Delta$ [that is given in the BdG Hamiltonian \eqref{eq:spin-up-bdg-hamiltonian}] for the chiral $d$-wave state in the extended Kane-Mele model:
\begin{align}
    \Delta_\alpha(\vv{q};\vv{k}) = \Delta_\alpha^{E_2}(\vv{q}) \klr{\psi_{x^2-y^2}^{E_2}(\vv{k}) + i \psi_{xy}^{E_2}(\vv{k})}  \,,
    \label{eq:ansatz-gap-function-chiral-d}
\end{align}
where $\Delta_\alpha^{E_2}(\vv{q}) \in \mathbb{R}$ represents the $\vv{k}$-independent order parameter of the system. 
In Fig.~\ref{fig:dwave-basis-functions} we have plotted the real and imaginary parts of the chiral $d$-wave superconducting gap function. In general, its value depends on the band~$\alpha$, i.e., we need to solve a two-dimensional fix-point equation here
\begin{align}
    \Delta^{E_2}_\alpha(\vv{q}) \nonumber
    =& \frac{U_0}{V} \!\sum_{\vv{k},n} \bar{\psi}^{E_2}_{x^2-y^2}(\vv{k}) v_{n\alpha,+}(\vv{q};\vv{k}) v^\ast_{n\alpha,-}(\vv{q};\vv{k})\\
    &\quad\quad\times\klr{1 \!-\! n_{\mathrm{F}}(E_{\vv{k}n}(\vv{q}))} , \label{eq:self-consistent-mean-field-dwave1}
\end{align}
for $\alpha = 1,2$. As a consistency check, we also verified that
\begin{align}
i\Delta^{E_2}_\alpha(\vv{q}) \nonumber
    =& \frac{U_0}{V} \!\sum_{\vv{k},n} \bar{\psi}^{E_2}_{xy}(\vv{k}) v_{n\alpha,+}(\vv{q};\vv{k}) v^\ast_{n\alpha,-}(\vv{q};\vv{k})\\
    &\quad\quad\times\klr{1 \!-\! n_{\mathrm{F}}(E_{\vv{k}n}(\vv{q}))}\,
\end{align}
is fulfilled when inserting the solution of Eq.~\eqref{eq:self-consistent-mean-field-dwave1}. For $M = 0$, the gap function is proportional to the identity matrix such that we only need to solve a two-dimensional fix-point equation as the order parameters are independent of the band index~$\alpha$. For example, in Fig.~\ref{fig:self-consistent-d-wave} of Appendix~\ref{sec:appendix-fig-s-wave} we have plotted for various interactions and fixed hopping phase~$\varphi = 1.377$ the self-consistent solutions we have obtained.

Compared to conventional $s$-wave superconductivity in Fig.~\ref{fig:superfluid-weight-along-M-axis-s-wave}, a chiral $d$-wave superconducting state seems to decrease the total superfluid weight by approximately $0.15t$ near $M = 0$ (cf.\ Fig.~\ref{fig:superfluid-weight-along-M-axis-d-wave}). However, as we increase $M$ and move over to the topological trivial sector, the total superfluid weight increases as the dominating conventional contribution acquires higher values which is possibly due to the $\vv{k}$ dependency in the $d$-wave pairing potential. In addition, the functional superfluid weight seems to play a negligible role as it is of order $0.0015t$. Nevertheless, this contribution is nonzero and may be significantly larger in other models. \\
\indent In Fig.~\ref{fig:func-weight-dwave-plane} we have plotted the functional superfluid weight with respect to the hopping phase~$\varphi$ and the staggered onsite potential~$M$. Even though the orbitals are fixed to the maximal Wyckoff positions, a nonzero functional superfluid weight appears in certain regions of the diagram. This suggests that the concept of the minimal quantum metric needs to be extended such that it also captures the behavior of the superfluid weight for unconventional superconducting states. As this contribution contains derivatives of the single-particle Hamiltonian, it is expected that the underlying mechanism is due to quantum geometry. The geometrical superfluid weight admits a lower bound determined by the Chern number which is also reflected in Fig.~\ref{fig:geom-weight-dwave-plane} where we observe an increased value for the geometrical superfluid weight in areas where nontrivial topology is present. It becomes evident in Fig.~\ref{fig:func-weight-dwave-plane} that the mechanism behind the functional contribution needs to be similar. \\
\indent In Sec.~\ref{sec:isolated-band-limit} we have derived an expression for the superfluid weight within the isolated narrow-bands limit with the help of perturbation theory. In particular, the calculations indicate that the underlying mechanism behind the functional contribution needs to be of different quantum-geometrical nature: We have shown that another quantum-geometrical quantity, such as a nontrivial Wilczek-Zee connection \eqref{eq:wilczek-zee-connection-definition} or the two-point fidelity magnitude \eqref{eq:two-point-fidelity-tensor-definition}, may explain the behavior of the functional superfluid weight. Indeed, for example, Eq.~\eqref{eq:qgt-radial-component} shows that, in two-band systems, the two-point fidelity magnitude, evaluated for $\vv{k} = \vv{k}'$ and $n'=n$, can be interpreted as a measure of the \enquote{radial component} of the quantum geometric tensor (QGT):
\begin{align}
    \zeta_{ij}^{(nn)}(\vv{k},\vv{k}) = |Q_{ij}^{(n)}(\vv{k})| \,.
\end{align}
Along the diagonal the Berry curvature is zero such that $\zeta_{ii}^{(nn)}(\vv{k},\vv{k}) = g_{ii}^{(n)}(\vv{k})$. Consequently, similar to the geometrical superfluid weight, one expects the functional contribution to be amplified in topologically nontrivial regions as well. But, note that the functional superfluid weight is additionally amplified due to the $\vv{k} \neq \vv{k}'$ summation in which the diagonal components of the two-point fidelity magnitude do not reduce to single-momentum quantum geometry. This behavior agrees well with Fig.~\ref{fig:func-weight-dwave-plane}. Unfortunately, for $M \neq 0$ the gap function is not proportional to the identity matrix, thus, it is needed to find a suitable model where the considerations done in Sec.~\ref{sec:isolated-band-limit} are applicable and can be analyzed in more detail. 

\section{Summary and outlook}

The main goal of this work was to derive an expression for the superfluid weight of mean-field BCS theory for systems with time-reversal symmetry and arbitrary unconventional pairing mechanism at zero temperature. 
While the conventional and geometrical contributions given in Eqs.~\eqref{eq:superfluid-weight-conv-def} and \eqref{eq:superfluid-weight-geom} are not altered for unconventional superconducting states, the key difference here is that the gap function becomes $\vv{k}$ dependent and, thus, derivatives with respect to the gap function need to be treated as functional derivatives. 
The result is a generalization of the expression for the superfluid weight in Ref.~\cite{huhtinen2022revisiting} that has been elaborated in Sec.~\ref{sec:calculation-superfluid-weight-functional}.
The obtained \enquote{functional superfluid weight} originating from the functional dependence of the free energy contains additional information about the superconducting state which is in general not reducible to a single-momentum quantum metric.

\indent We have derived an expression [cf.\ Eq.~\eqref{eq:functional-superfluid-weight-wrt-skew-matrices}] for the functional superfluid weight that can be easily implemented numerically and enhanced the discussion of the isolated band limit for bands that are not flat with the help of quantum-mechanical perturbation theory in Sec.~\ref{sec:isolated-band-limit}. 
In particular, we obtained expressions for the conventional, geometrical, and functional superfluid weight within the isolated-band limit at zeroth order in perturbation theory. 
It is well known that the geometrical contribution is driven by a nonzero quantum metric. 
The functional superfluid weight, however, is rather driven by a nontrivial Wilczek-Zee connection and we have shown that it can be bounded by the two-point fidelity magnitude~\eqref{eq:two-point-fidelity-tensor-definition} in certain cases. 
The two-point fidelity magnitude is a quantity that is also related to quantum geometry which measures the \enquote{similarity} of orthogonal states after a small perturbation in parameter space.
It would be important to include higher-order corrections to the expressions obtained via perturbation theory to reproduce and extend the calculations that have been done in Ref.~\cite{jiang2024geometric}. \\
\indent Furthermore, we have explored our main results with the help of a specific model, employing the extended Kane-Mele Hamiltonian introduced in Ref.~\cite{lau2022universal}.
We benchmarked our numerical calculations for the $s$-wave case and implemented chiral $d$-wave superconductivity leading to the main result presented in Fig.~\ref{fig:func-weight-dwave-plane}: 
Similar to the geometrical superfluid weight, we observe that topology seems to influence the functional superfluid weight as well, as it reaches its maximum in the topologically nontrivial phase. 
Thus, it is reasonable to assume that this quantity is related to quantum geometry. 
In particular, within the isolated band limit, we have shown that quantum-geometrical quantities, such as a nontrivial Wilczek-Zee connection or the two-point fidelity magnitude, may explain the behavior of the functional superfluid weight. Since the two-point fidelity magnitude can be regarded as the radial component of the QGT, one can expect the functional contribution to be amplified for topological nontrivial systems as well.
However, for $M \neq 0$, the gap function is not proportional to the identity matrix, and we are therefore unable to compare our numerical results with Eq.~\eqref{eq:explicitfunc} directly. 
More investigation and a different single-particle model is needed to determine in more detail in what extent the considerations presented in Sec.~\ref{sec:isolated-band-limit} are applicable. \\
\indent In this work, we have defined the superfluid weight as the second total derivative of the free energy with respect to the gauge fields which is equivalent to the definition that is widely used in the literature based on linear response theory. For example, in Ref.~\cite{lamponen2025superconductivity} the authors have derived similar expressions for the superfluid weight for systems with nearest-neighbor pairing. A more detailed analysis and a direct comparison is necessary to evaluate whether both expressions are identical.
Moreover, we believe it would be instructive to reverse the order of the calculation steps that have been done in Sec.~\ref{sec:bcs-theory}, as it should not matter if one diagonalizes the single-particle Hamiltonian in the BCS partition function~\eqref{eq:bcs-partition-function} first and then perform a saddle-point Hubbard-Stratonovich transformation. 
The formulas may change slightly but the quantum geometric features  should stay the same. 
It is also important to go beyond mean-field theory and to quantify the corrections that appear due to the fluctuations around the saddle-point solution. Additionally, an incorporation of disorder (e.g., as stochastic noise) is also necessary to analyze, in what extent it influences the superfluid weight of a system. 
Lastly, it would be also interesting to take a look at TRS breaking systems or systems exhibiting interband pairing, and generalize the equations for the superfluid weight accordingly.

\begin{acknowledgments}
We thank K.-E.~Huhtinen and B.~Hawashin for helpful discussions. Moreover, we thank E.\ Lamponen, S.~P\"ontys, and P.~T\"orm\"a for correspondence and comments on our manuscript.
M.M.S. acknowledges funding from the Deutsche Forschungsgemeinschaft (DFG, German Research Foundation) under Project No.~277146847 (SFB 1238, Project No.~C02) and Project No.~452976698 (Heisenberg program).
T.H. acknowledges financial support by the 
European Research Council (ERC) under grant QuantumCUSP
(Grant Agreement No.~101077020).
\end{acknowledgments}

\section*{Data availability}
The data that support the findings of this article are openly available \cite{dataset}.

\appendix

\section{Hellmann-Feynman theorem}\label{app:hellmann-feynman}
Suppose you have some parameter-dependent non-degenerate Hamiltonian $H(\vv{k})$ where $\vv{k}$ denotes some $N$-dimensional parameter. Then, the eigenvalues $E_n(\vv{k})$ are also parameter dependent. The Hellmann-Feynman theorem states that it is possible to express derivatives of the eigenvalues $E_n(\vv{k})$ in terms of derivatives of the underlying Hamiltonian~\cite{hellmann1937einfuhrung,feynman1939forces}. So, suppose $U \equiv U(\vv{k})$ is the unitary matrix that diagonalizes the Hamiltonian $H \equiv H(\vv{k})$, i.e., the property
\begin{align}
    U^\dagger_{n\alpha} H_{\alpha\beta} U_{\beta n} = E_n
    \label{eq:U-diagonalizes-H}
\end{align}
holds, and, since $U$ is unitary we also have
\begin{align}
    U_{\alpha n}^\dagger U_{n \beta} = \delta_{\alpha\beta} \,.
    \label{eq:unitarity-of-U}
\end{align}
\paragraph{First-order derivative.} If we calculate the first derivative of $E_n$ with respect to $k_i$ we obtain by using the product rule
\begin{align}
    \partial_i E_n =& U^\dagger_{n\alpha} \partial_i H_{\alpha\beta} U_{\beta n} + \partial_i U^\dagger_{n\gamma} \delta_{\gamma\alpha} H_{\alpha\beta} U_{\beta n} \nonumber \\
    &+ U^\dagger_{n\alpha} H_{\alpha\beta} \delta_{\beta\gamma} \partial_i U_{\gamma n} \,.
\end{align}
Replace the Kronecker deltas with Eq.~\eqref{eq:unitarity-of-U} and we end up with
\begin{align}
    \partial_i E_n &= U^\dagger_{n\alpha} \partial_i H_{\alpha\beta} U_{\beta n} \,.
\end{align}
\paragraph{Second-order derivative.} According to the product rule, the second derivative of the $n$th eigenvalue is given by
\begin{align}
    \partial_i \partial_j E_n =& U^\dagger_{n\alpha} \partial_i \partial_j H_{\alpha\beta} U_{\beta n} + \partial_i U^\dagger_{n\alpha} \partial_j H_{\alpha\beta} U_{\beta n} \nonumber \\
    &+ U^\dagger_{n\alpha} \partial_j H_{\alpha\beta} \partial_i U_{\beta n} \,.
    \label{eq:int-result-hellman-feynman}
\end{align}
Due to Eq.~\eqref{eq:U-diagonalizes-H} we have for $n \neq m$ the following property:
\begin{align}
    U^\dagger_{n\alpha} H_{\alpha\beta} U_{\beta m} = 0 \,.
\end{align}
If we take the derivative on both sides and make use of $\partial_i U^\dagger U = - U^\dagger \partial_i U$, it leads us to the following expression:
\begin{align}
    U^\dagger_{n\alpha} \partial_i U_{\alpha m} = \frac{U^\dagger_{n\alpha} \partial_i H_{\alpha\beta} U_{\beta m}}{E_m - E_n} \,.
\end{align}
We insert this result into Eq.~\eqref{eq:int-result-hellman-feynman} and obtain
\begin{align}
    \partial_i \partial_j E_n =& \kle{U^\dagger \partial_i \partial_j H U}_{n,n} \nonumber \\
    &+ \sum_{m \neq n} \klr{\frac{\kle{U^\dagger \partial_i H U}_{n,m} \kle{U^\dagger \partial_j H U}_{m,n}}{E_n - E_m} + (i \leftrightarrow j)} \,.
    \label{eq:hellman-feynman-2nd-derivative}
\end{align}
This theorem is also valid if one takes functional derivatives of the eigenvalues since one can express a functional derivative as a partial derivative. 

\section{Useful matrix summation}\label{sec:summation}

We need to calculate sums of the form
\begin{widetext}
\begin{align}
    s(Y_1,Y_2) = \sum_{n=1}^{2N_{\mathrm{B}}} \sum_{\substack{m=1\\m \neq n}}^{2N_{\mathrm{B}}} \frac{\mathrm{sgn}(E_{\vv{k}n})}{E_{\vv{k}n} - E_{\vv{k}m}} \kle{\matr{X_1 & Y_1\\ Y_1^\dagger & Z_1}_{nm} \matr{X_2 & Y_2\\ Y_2^\dagger & Z_2}_{mn} + \matr{X_2 & Y_2\\ Y_2^\dagger & Z_2}_{nm} \matr{X_1 & Y_1\\ Y_1^\dagger & Z_1}_{mn}} \,,
    \label{eq:definition-sum-s}
\end{align}
where $X_{1,2},Y_{1,2},Z_{1,2}$ denote complex-valued ($N_{\mathrm{B}} \times N_{\mathrm{B}}$) matrices. If we perform a similar calculation to Ref.~\cite{xie2020topology}, it turns out that $s$ depends on the off-diagonal block matrices $Y_1$ and $Y_2$ only. However, as soon as we consider a nonzero temperature, the $\mathrm{sgn}$ function is replaced by Fermi-Dirac functions and the other matrices become also important. We expand Eq.~\eqref{eq:definition-sum-s} to obtain
\begin{align}
    s(Y_1,Y_2) =& \sum_{n=1}^{N_{\mathrm{B}}} \sum_{\substack{m=1\\m\neq n}}^{N_{\mathrm{B}}} \kle{\frac{\kle{X_1}_{nm} \kle{X_2}_{mn}}{E_{\vv{k}n} - E_{\vv{k}m}} + \frac{\kle{X_2}_{nm} \kle{X_1}_{mn}}{E_{\vv{k}n} - E_{\vv{k}m}} - \frac{\kle{Z_1}_{nm} \kle{Z_2}_{mn}}{E_{\vv{k}(n+N_{\mathrm{B}})} - E_{\vv{k}(m+N_{\mathrm{B}})}} - \frac{\kle{Z_2}_{nm} \kle{Z_1}_{mn}}{E_{\vv{k}(n+N_{\mathrm{B}})} - E_{\vv{k}(m+N_{\mathrm{B}})}}} \\
    &+ \sum_{n,m=1}^{N_{\mathrm{B}}} \kle{\frac{\kle{Y_1}_{nm} [Y_2^\dagger]_{mn}}{E_{\vv{k}n} - E_{\vv{k}(m+N_{\mathrm{B}})}} + \frac{\kle{Y_2}_{nm} [Y_1^\dagger]_{mn}}{E_{\vv{k}n} - E_{\vv{k}(m+N_{\mathrm{B}})}} - \frac{[Y_1^\dagger]_{nm} \kle{Y_2}_{mn}}{E_{\vv{k}(n+N_{\mathrm{B}})} - E_{\vv{k}m}} - \frac{[Y_2^\dagger]_{nm} \kle{Y_1}_{mn}}{E_{\vv{k}(n+N_{\mathrm{B}})} - E_{\vv{k}m}}} \,.
\end{align}
We relabel $(m \leftrightarrow n)$ in the first, third, fifth, and sixth terms and change the order of the factors to get
\begin{align}
    s(Y_1,Y_2) =& \sum_{n=1}^{N_{\mathrm{B}}} \sum_{\substack{m=1\\m\neq n}}^{N_{\mathrm{B}}} \kle{\frac{\kle{X_2}_{nm} \kle{X_1}_{mn}}{E_{\vv{k}m} - E_{\vv{k}n}} + \frac{\kle{X_2}_{nm} \kle{X_1}_{mn}}{E_{\vv{k}n} - E_{\vv{k}m}} - \frac{\kle{Z_2}_{nm} \kle{Z_1}_{mn}}{E_{\vv{k}(m+N_{\mathrm{B}})} - E_{\vv{k}(n+N_{\mathrm{B}})}} - \frac{\kle{Z_2}_{nm} \kle{Z_1}_{mn}}{E_{\vv{k}(n+N_{\mathrm{B}})} - E_{\vv{k}(m+N_{\mathrm{B}})}}} \label{eq:vanishing-line-calc} \\
    &+ \sum_{n,m=1}^{N_{\mathrm{B}}} \kle{\frac{[Y_2^\dagger]_{nm} [Y_1]_{mn}}{E_{\vv{k}m} - E_{\vv{k}(n+N_{\mathrm{B}})}} + \frac{[Y_1^\dagger]_{nm} [Y_2]_{mn}}{E_{\vv{k}m} - E_{\vv{k}(n+N_{\mathrm{B}})}} - \frac{[Y_1^\dagger]_{nm} \kle{Y_2}_{mn}}{E_{\vv{k}(n+N_{\mathrm{B}})} - E_{\vv{k}m}} - \frac{[Y_2^\dagger]_{nm} \kle{Y_1}_{mn}}{E_{\vv{k}(n+N_{\mathrm{B}})} - E_{\vv{k}m}}} \label{eq:non-vanishing-line-calc} \,,
\end{align}
i.e., all terms in Eq.~\eqref{eq:vanishing-line-calc} cancel each other out, we can group two terms in Eq.~\eqref{eq:non-vanishing-line-calc} and we are left with
\begin{align}
     s(Y_1,Y_2) &= 2\sum_{n,m=1}^{N_{\mathrm{B}}} \kle{\frac{\kle{Y_1}_{nm} [Y_2^\dagger]_{mn}}{E_{\vv{k}n} + E_{-\vv{k}m}} + \frac{\kle{Y_2}_{nm} [Y_1^\dagger]_{mn}}{E_{\vv{k}n} + E_{-\vv{k}m}}} = \sum_{n,m=1}^{N_{\mathrm{B}}} \frac{4\, \mathrm{Re}[(Y_1)_{nm} (Y_2)_{nm}^\ast]}{E_{\vv{k}n} + E_{-\vv{k}m}}\,,
     \label{eq:important-sum-fh}
\end{align}
\end{widetext}
whereby we used the relation $E_{\vv{k}n} = -E_{-\vv{k}(n+N_{\mathrm{B}})}$.
\section{Sherman-Morrison-Woodbury formula}\label{sec:woodbury}
With the help of the Sherman-Morrison-Woodbury formula we can calculate the inverse of the matrix provided in Eq.~\eqref{eq:definition-matrix-M}:
\begin{align}
    \hat{M} = V\hat{U}^{-1} - \hat{\Pi} = V\hat{U}^{-1} \klr{\hat{\Pi}^{-1} - \frac{1}{V}\hat{U}} \hat{\Pi}
\end{align}
exactly in case the interaction factorizes such that there exist functions $u_1(\vv{k}),\hdots,u_d(\vv{k})$ with
\begin{align}
    U(\vv{k},\vv{k}') = U_0 \sum_{i=1}^d u_i(\vv{k}) u_i(\vv{k}') \,.
\end{align}
If this is the case, we can calculate the inverse of $\hat{M}$ by interpreting $\hat{U}/V$ as a rank-$2dN_{\mathrm{B}}$ update to the inverse of $\hat{\Pi}^{-1}$. For example, if a conventional $s$-wave pairing mechanism is in our system present, it corresponds to a superconducting state associated with a one-dimensional irrep, i.e., we can interpret $\hat{U}$ as a rank-$2N_{\mathrm{B}}$ update. Another example occurs in our numerical calculations where we consider chiral $d$-wave superconductivity on a hexagonal lattice. In this case the functions $u_i(\vv{k})$ represent the form factors, i.e., this situation translates to a rank-$4N_{\mathrm{B}}$ update. \\
\indent Let us define the vectors
\begin{align}
    \ket{x_{i\alpha\mu}} \coloneqq -\frac{U_0}{V} \sum_{\vv{k}} u_i(\vv{k}) \ket{\vv{k}\alpha\mu}\,, \quad 
    \ket{y_{i\alpha\mu}} \coloneqq \sum_{\vv{k}} u_i(\vv{k}) \ket{\vv{k}\alpha\mu}
\end{align}
and the matrices
\begin{align}
    \acute{L} &\coloneqq \kle{\ket{x_{1\mathrm{R}1}}, \hdots, \ket{x_{1\mathrm{I}N_{\mathrm{B}}}},\ket{x_{2\mathrm{R}1}},\hdots,\ket{x_{2\mathrm{I}N_{\mathrm{B}}}}}\,,\\
    \grave{R} &\coloneqq \kle{\bra{y_{1\mathrm{R}1}}, \hdots, \bra{y_{1\mathrm{I}N_{\mathrm{B}}}},\bra{y_{2\mathrm{R}1}},\hdots,\bra{y_{2\mathrm{I}N_{\mathrm{B}}}}}^T \,,
\end{align}
such that $\hat{U} = \acute{L}\grave{R}$. According to the Sherman-Morrison-Woodbury formula~\cite{woodbury1950inverting}, the inverse of $\hat{M}$ is given by
\begin{align}
    \hat{M}^{-1} &= -\hat{\Pi}^{-1} \klr{\hat{\Pi}^{-1} + \acute{L}\grave{R}}^{-1} \acute{L}\grave{R} \\
    &= -\hat{\Pi}^{-1}\klr{\hat{\Pi} - \hat{\Pi} \acute{L}\klr{\check{\mathbbm{1}} + \grave{R}\hat{\Pi}\acute{L}}^{-1} \grave{R} \hat{\Pi}} \acute{L}\grave{R}\,,
\end{align}
where $\check{\mathbbm{1}}$ denotes the ($2dN_{\mathrm{B}} \times 2dN_{\mathrm{B}}$) identity matrix. We define the components of the ($2dN_{\mathrm{B}} \times 2dN_{\mathrm{B}}$) matrix $\check{\Sigma}$ by
\begin{align}
    \Sigma_{i\alpha\mu,l\beta\nu} &\coloneqq \bra{y_{i\alpha\mu}}\hat{\Pi}\ket{x_{l\beta\nu}} = -\frac{U_0}{V} \sum_{\vv{k}} u_i(\vv{k}) u_l(\vv{k}) \Pi_{\alpha\mu,\beta\nu}(\vv{k}) \,.
\end{align}
Set $\ket{u_{i\alpha\mu}} = \sum_{\vv{k}} u_i(\vv{k}) \ket{\vv{k}\alpha\mu}$ and it turns out that the inverse of $\hat{M}$ is given by
\begin{align}
    \hat{M}^{-1} &= \frac{U_0}{V} \!\! \sum_{i,l=1}^d \sum_{\alpha,\beta = 1}^{N_{\mathrm{B}}} \sum_{\mu,\nu = \mathrm{R},\mathrm{I}} \ket{u_{i\alpha\mu}} \big[(\check{\mathbbm{1}} + \check{\Sigma})^{-1}\big]_{i\alpha\mu,l\beta\nu} \bra{u_{l\beta\nu}} \,,
\end{align} 
hence its components can be expressed as
\begin{align}
    M^{-1}_{\alpha\mu,\beta\nu}(\vv{k},\vv{k}') = \frac{U_0}{V} \sum_{i,l=1}^d u_i(\vv{k}) u_l(\vv{k}') \big[(\check{\mathbbm{1}} + \check{\Sigma})^{-1}\big]_{i\alpha\mu,l\beta\nu} \,.
    \label{eq:unconventional-M}
\end{align}
If we have a conventional pairing mechanism with a TRS single-particle Hamiltonian, then $u_i(\vv{k}) = 1$, $d = 1$, and $\mu=\nu=\mathrm{I}$ (we can drop the indices $i,l,\mu,\nu$) such that the expression reduces to
\begin{align}
    M^{-1}_{\alpha\beta} = \frac{U_0}{V}  \big[(\check{\mathbbm{1}} + \check{\Sigma})^{-1}\big]_{\alpha\beta}\,,
    \label{eq:conventional-M}
\end{align}
which is independent of $\vv{k},\vv{k}'$.

\section{Wilczek-Zee connection and fidelity}\label{app:wilczek-fidelity}
The Wilczek-Zee connection, introduced by Wilczek and Zee in 1984 \cite{wilczek1984appearance}, naturally generalizes the Berry connection in the sense that one considers the projection of the initial state $\ket{\psi_n(\vv{k})}$ after a small variation of parameters onto a different state $\ket{\psi_m(\vv{k})}$. It is defined as
\begin{align}
    e_{i,m}^{(n)}(\vv{k}) \coloneqq i \braket{\psi_m(\vv{k})}{\partial_i \psi_n(\vv{k})}\,,
    \label{eq:wilczek-zee-connection-definition-app}
\end{align}
whereby we have adapted the notation by Romero \textit{et al.}\ \cite{romero2024n}. While this quantity has been already discussed in the context of nuclear quadrupole resonance \cite{tycko1987adiabatic,zee1988non}, initial ideas date back to a work by Herzberg and Teller in 1933 in theoretical chemistry \cite{herzberg1933schwingungsstruktur}, in which the Wilczek-Zee connection is nowadays known by the name nonadiabatic coupling vector~\cite{yarkony2002nonadiabatic}. In this work it appears in Sec.~\ref{sec:isolated-band-limit} within the functional contribution to the superfluid weight; see also Ref.~\cite{yijian2025quantum} where the Wilczek-Zee connection has been discussed within the geometrical contribution of the superfluid weight. \\
\indent A nonzero Berry connection can be interpreted as the ability of the state to remain in the same state after a small variation of parameters. Hence, analogously, since a nonzero Wilczek-Zee connection $e_{i,m}^{(n)}$ implies a non-zero probability to go from the $n$th state to the $m$th state, it can be interpreted as the inability of the state to remain in the same state after a small variation of the parameters \cite{romero2024n}. Mathematically speaking, a zero Wilczek-Zee connection $e_{i,m}^{(n)} = 0$ (for all $m$) implies that $\kls{\braket{\psi_n(\vv{k})}{\psi_n(\vv{k}+\mathrm{d}\vv{k})}} = 1$ for arbitrary small variations $\mathrm{d}\vv{k}$. In this case, the state $\ket{\psi_n(\vv{k})}$ is stationary and does not move in the projective Hilbert space, i.e., we can conclude that a trivial Wilczek-Zee connection implies trivial quantum geometry. \\
\indent Due to the Hellman-Feynman theorem (cf.\ Appendix \ref{app:hellmann-feynman}), the Wilczek-Zee connection admits an expression that is similar to the Zanardi-Giorda-Cozzini representation \cite{zanardi2007information} of the quantum geometric tensor (QGT)
\begin{align}
    Q_{ij}^{(n)} = \sum_{m \neq n} \frac{\bra{\psi_n}\partial_i H\ket{\psi_m}\bra{\psi_m}\partial_j H\ket{\psi_n}}{\klr{\varepsilon_n - \varepsilon_m}^2} \,,
    \label{eq:zanardi-giorda-cozzini}
\end{align}
which is given by \cite{romero2024n}
\begin{align}
    e_{i,m}^{(n)}(\vv{k}) = i\frac{\bra{\psi_m(\vv{k})}\partial_i H(\vv{k})\ket{\psi_n(\vv{k})}}{\varepsilon_n(\vv{k}) - \varepsilon_m(\vv{k})} \,.
\end{align}
Hence, it is possible to represent the QGT in terms of the Wilczek-Zee connection. Inserting this relation into Eq.~\eqref{eq:zanardi-giorda-cozzini} we obtain the Wilczek-Zee representation of the QGT:
\begin{align}
    Q_{ij}^{(n)}(\vv{k}) = \sum_{m \neq n} \bar{e}_{i,m}^{(n)}(\vv{k})\, e_{j,m}^{(n)}(\vv{k}) \,,
    \label{eq:qgt-wilczek-zee-representation}
\end{align}
where the bar indicates complex conjugation. Equivalent representations for the quantum metric and the Berry curvature can be found in Ref.~\cite{romero2024n}. \\
\indent A related quantity is the fidelity between two (normalized) states $\ket{\psi(\vv{k})}$ and $\ket{\phi(\vv{k})}$ which is defined as~\cite{jozsa1994fidelity}
\begin{align}
    F(\ket{\psi(\vv{k})},\ket{\phi(\vv{k})}) \coloneqq \kls{\braket{\psi(\vv{k})}{\phi(\vv{k})}} \,.
\end{align}
Consider a small variation and perform a Taylor expansion
\begin{align}
    &F(\ket{\psi(\vv{k})},\ket{\phi(\vv{k} + \mathrm{d}\vv{k})}) 
    \nonumber \\
    &\quad= \kls{\braket{\psi(\vv{k})}{\phi(\vv{k})} + \braket{\psi(\vv{k})}{\partial_i \phi(\vv{k})} \mathrm{d}k_i + \hdots} \,.
\end{align}
In particular, if the states $\ket{\psi(\vv{k})}$ and $\ket{\phi(\vv{k})}$ are orthogonal to each other, the expression simplifies (up to first order) to
\begin{align}
    F(\ket{\psi(\vv{k})},\ket{\phi(\vv{k} + \mathrm{d}\vv{k})}) = \kls{\braket{\psi(\vv{k})}{\partial_i \phi(\vv{k})}} \mathrm{d}k_i \,.
\end{align}
Suppose that $\ket{\psi(\vv{k})}$ and $\ket{\phi(\vv{k})}$ represent the $n$th and $m$th eigenstates of the Hamiltonian and consider the total fidelity between the $n$th state and all the other states, i.e., 
\begin{align}
    \sum_{m \neq n} F(\ket{\psi_m(\vv{k})},\ket{\psi_n(\vv{k} + \mathrm{d}\vv{k})}) = \sum_{m \neq n} \kls{\braket{\psi_m(\vv{k})}{\partial_i \psi_n(\vv{k})}} \mathrm{d}k_i \,.
\end{align}
We define the resulting quantity as the one-point fidelity magnitude $\zeta_i^{(n)}(\vv{k})$,
\begin{align}
    \zeta_i^{(n)}(\vv{k}) \coloneqq \sum_{m \neq n} \kls{\braket{\psi_m(\vv{k})}{\partial_i \psi_n(\vv{k})}} \,,
\end{align}
which measures the similarity of the $n$th band to the other bands after a small variation of the parameters. \\
\indent We can proceed analogously and define the two-point fidelity between two states $\ket{\psi(\vv{k})}$ and $\ket{\phi(\vv{k})}$ at two different points $\vv{k},\vv{k}'$ in the Brillouin zone as the product of two one-point fidelities:
\begin{align}
    &F_{\mathrm{two-point}}(\ket{\psi(\vv{k})},\ket{\phi(\vv{k})};\ket{\psi(\vv{k}')},\ket{\phi(\vv{k}')}) \nonumber \\
    &\quad \coloneqq F(\ket{\psi(\vv{k})},\ket{\phi(\vv{k})}) \cdot F(\ket{\psi(\vv{k}')},\ket{\phi(\vv{k}')}) \,.
\end{align}
If we perform a small variation and proceed as above, we end up with the two-point fidelity magnitude~$\zeta_{ij}^{(nn')}(\vv{k},\vv{k}')$:
\begin{align}
    \zeta_{ij}^{(nn')}(\vv{k},\vv{k}') \coloneqq \sum_{\mb{m\neq n,m' \neq n'}} \kls{\braket{\psi_m(\vv{k})}{\partial_i \psi_n(\vv{k})} \braket{\psi_{m'}(\vv{k}')}{\partial_j \psi_{n'}(\vv{k}')}} \,.
\end{align}
We identify the Wilczek-Zee connection \eqref{eq:wilczek-zee-connection-definition-app}
\begin{align}
    \zeta_{ij}^{(nn')}(\vv{k},\vv{k}') = \sum_{m\neq n,m' \neq n'} \kls{e_{i,m}^{(n)}(\vv{k}) e_{j,m'}^{(n')}(\vv{k}')}
\end{align}
and realize, that this quantity is quite similar to the QGT in Wilczek-Zee representation \eqref{eq:qgt-wilczek-zee-representation}. Moreover, for $\vv{k}' = \vv{k}$ and $n'=n$ it actually represents an upper bound of the QGT 
\begin{align}
    \zeta_{ij}^{(nn)}(\vv{k},\vv{k}) &= \zeta_i^{(n)}(\vv{k}) \zeta_j^{(n)}(\vv{k}) \nonumber \\
    &\ge \kls{Q_{ij}^{(n)}(\vv{k})} = \sqrt{\klr{g^{(n)}_{ij}}^2 + \frac{1}{4} \klr{F_{ij}^{(n)}}^2} \label{eq:qgt-radial-component}
\end{align}
due to the Cauchy-Schwarz inequality whereas the inequality is always saturated for two-band systems, i.e., the two-point fidelity magnitude can be regarded as a measure of the radial component of the QGT, in particular as the sum of the quantum metric and the Berry curvature. Hence, this quantity integrated admits also as a lower bound given by the Chern number. However, keep in mind that the two-point fidelity magnitude neither constitutes a metric nor a tensor; it is merely a gauge-invariant measure of the \enquote{similarity} between two orthogonal states after a small parameter variation. 

\onecolumngrid
\section{Figure for Secs.~\ref{sec:kane-mele-s-wave} and \ref{sec:kane-mele-d-wave}}\label{sec:appendix-fig-s-wave}


\begin{figure*}[h!]
    \centering
    \subfigure[]{
        \includegraphics[width=0.45\linewidth]{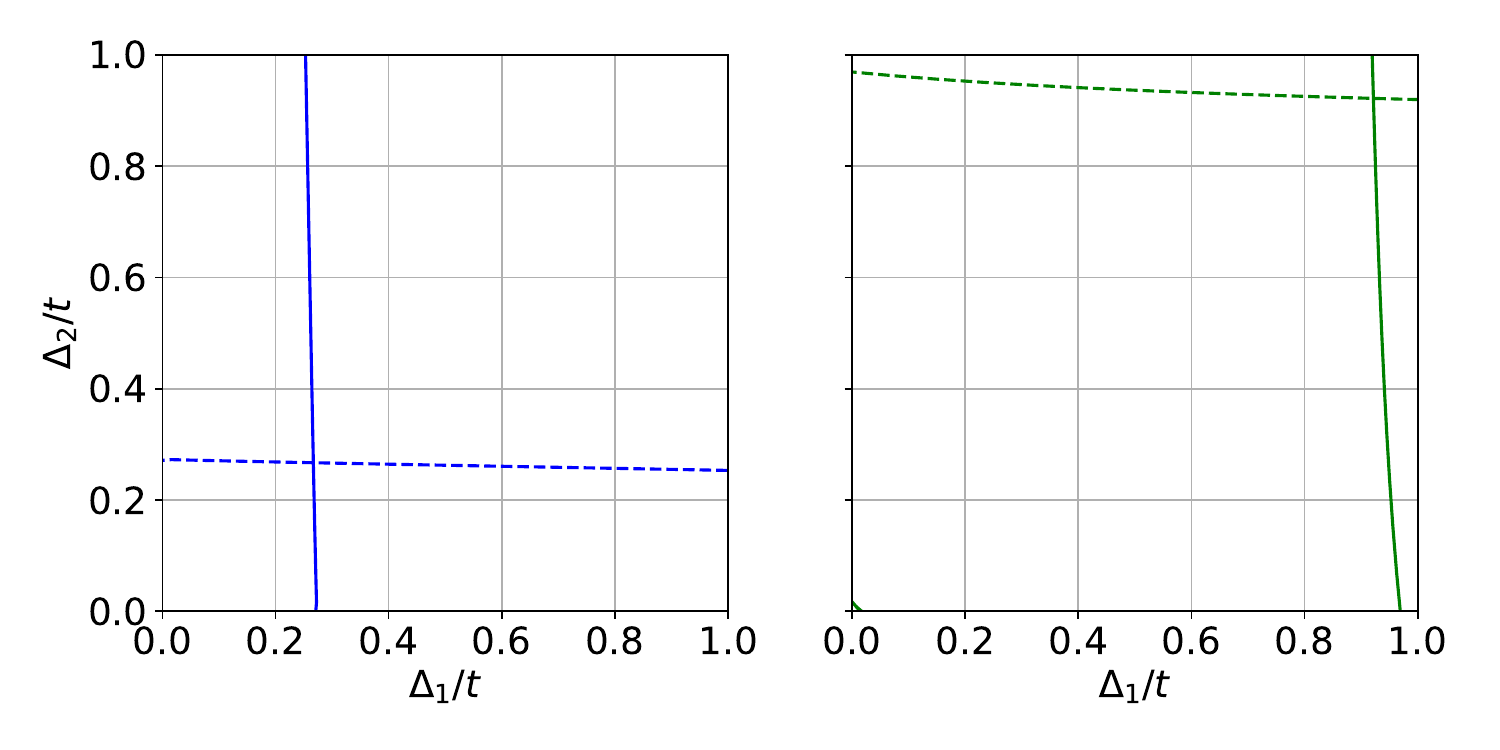}
        \label{fig:s-wave-self-consistent-equations}
    }
    \subfigure[]{
        \includegraphics[width=0.45\linewidth]{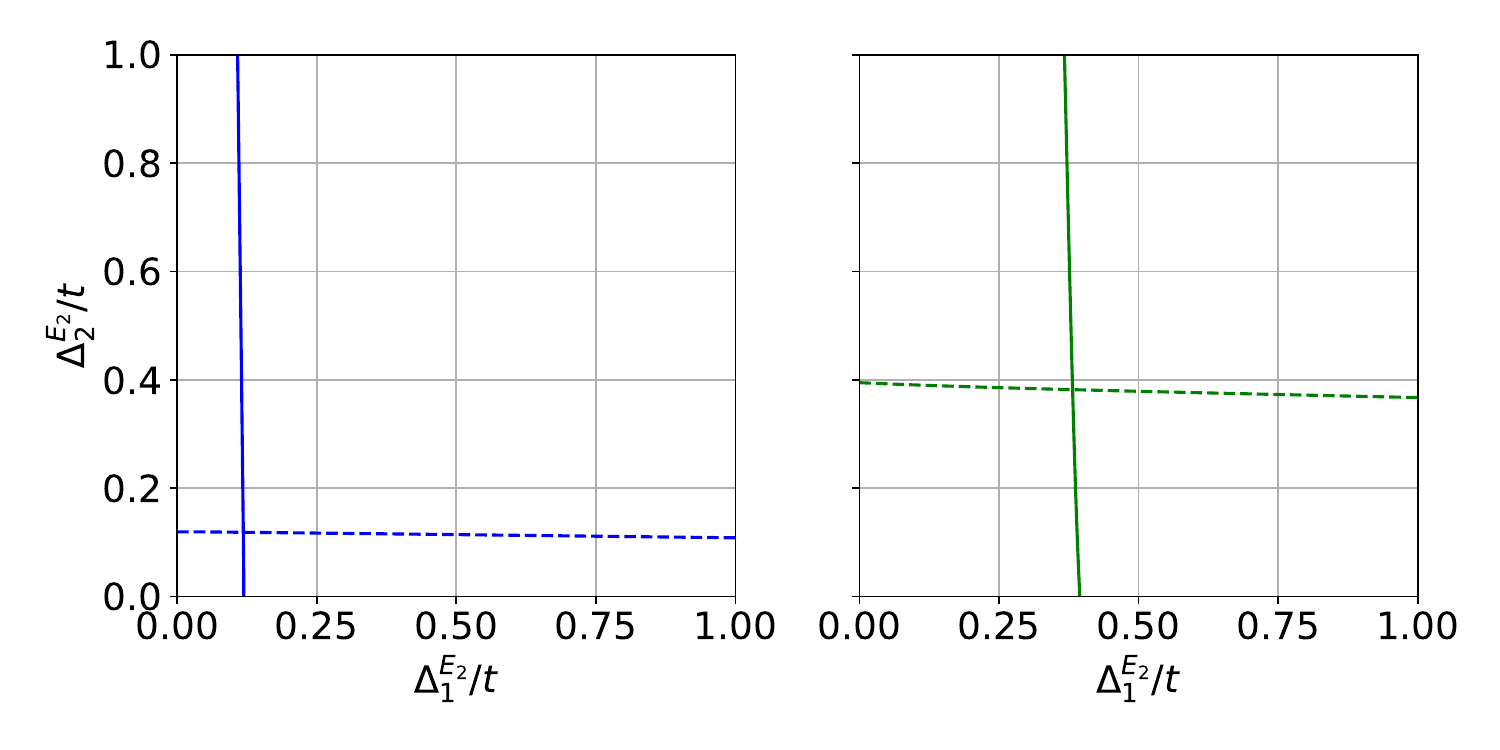}        \label{fig:self-consistent-d-wave}
    }
   \caption{Kane-Mele Hamiltonian for fixed $t_2 = 0.349t$, $t_3 = -0.264t$, $t_4 = 0.026t$, $M = 0$ and $\varphi = 1.377$ whereas the Brillouin zone has been discretized by a $20 \times 20$ grid. The numerical errors are of order $\le 1\%$. The two-dimensional self-consistent mean-field equations are recast as a fix-point equation of the form $\mathrm{func}_1(\Delta) = 0$ (dashed curve) and $\mathrm{func}_2(\Delta) = 0$ (solid curve), with model-parameters $M = 0$ and $\varphi = 1.377$ for various interaction strengths at zero temperature. \textbf{(a)} For the conventional superconducting state, we obtain $\Delta_1 = \Delta_2 = 0.267t$ for $U_0 = t$ (blue)
   and $\Delta_1 = \Delta_2 = 0.922t$ for $U_0 = 3t$ (green). \textbf{(b)} For the chiral $d$-wave superconducting state, we obtain ${\Delta_1}^{E_2} = {\Delta_2}^{E_2} = 0.118t$ for $U_0 = t$ (blue) 
   and ${\Delta_1}^{E_2} = {\Delta_2}^{E_2} = 0.382t$ for $U_0 = 3t$ (green).}
    \label{fig:conv-vs-geom-vs-func-s-wave}
\end{figure*} 
\twocolumngrid

\bibliography{apssamp}

\end{document}